\documentstyle[aps,epsfig]{revtex}

\begin{document}
\def\la{\langle}
\def\ra{\rangle}
\def\om{\omega}
\def\Om{\Omega}
\def\o0{\omega_0}
\def\ov{\omega_V}
\def\os{\omega_s}
\def\O0{\Omega_0}
\def\Os{\Omega_s}
\def\Om{\Omega}
\def\do{\Delta\omega}
\def\vep{\varepsilon}
\def\wh{\widehat}
\def\da{\dagger}
\def\tr{\rm{Tr}}
\newcommand{\beq}{\begin{equation}}
\newcommand{\eeq}{\end{equation}}
\newcommand{\beqa}{\begin{eqnarray}}
\newcommand{\eeqa}{\end{eqnarray}}
\newcommand{\intf}{\int_{-\infty}^\infty}
\newcommand{\into}{\int_0^\infty}
\title{Characteristic times in one dimensional scattering\footnote{To appear in ``Time in 
quantum mechanics'', Springer-Verlag; edited by J. G. Muga, R. Sala Mayato, and
I.L. Egusquiza}} 
\author{J. G. Muga}
\address{Departamento de Qu\'\i mica-F\'\i sica, Universidad del
Pa\'\i s Vasco, 
Apdo 644. Bilbao, Spain}

\maketitle              

\begin{abstract}
This chapter reviews several  
quantities that have been proposed in scattering theory   
to characterize the temporal aspects of one dimensional collisions: 
the dwell time, the delay time, the decay time, and times 
characterizing transient effects or the attainment of stationary 
conditions. Some aspects of tunnelling times are also discussed.  
\end{abstract}

\section{Introduction}
Quantum scattering theory deals with collisions, namely,
interactions which are 
essentially localized in time and space. 
This means that the interaction potential must vanish 
rapidly enough in coordinate space,
so that the wave packet tends to free-motion 
incoming and outgoing asymptotic states before and after the interaction is 
effective.   
The scope of scattering theory also includes ``half-collisions'' or  
``decay processes'' where the stage before the collision is ignored,
i.e., the evolution of the system is only considered from 
the interaction region. 

This chapter reviews various 
quantities that have been proposed in scattering theory   
to characterize the temporal aspects of the collision.  
A quantum wave packet collision with a potential
barrier in one dimension (1D)
is fully described by the evolution of the wave function $\psi(x,t)$
from the incoming to the outgoing asymptotic states. However,
the whole information contained in $\psi(x,t)$
is hardly required. A few well chosen  
quantities are often enough to provide a fair picture of the dynamics. 
In particular, one of these elementary parameters is
the transmission probability $P_T$, but to   
describe the time dependence  
we also need to quantify    
the duration of the collision, the arrival time at a detector,
the decay time of an unstable state, the asymptotic behaviour at short
and large times, or response times, such as the time required
to ``charge'' a well
or achieve stationary conditions when a source is turned on.

In spite of the inherent time dependence of collisions, 
the treatises on quantum mechanics or scattering theory concentrate
on solutions of the time-independent Schr\"odinger equation.
This is in part because many scattering experiments to obtain 
cross sections are performed in quasi-stationary
conditions, and also because the stationary scattering states
form a basis to analyze the actual time dependent collision.
In many cases wave packet scattering
is relegated to justify the
cavalier obtention by stationary methods of 
cross section expressions, and occasionally to discuss
resonance lifetimes. 
Another widespread limitation is the exclusive interest in 
the final results of the collision at asymptotic distances and times, 
which has been generally justified because   
``the midst of the collision cannot be observed''.  
However, while it is true that in many collision experiments   
only the asymptotic results are observed, 
modern experiments with femptosecond laser pulses 
or other techniques known as ``spectroscopy of the transition 
state'' do probe the structure and the evolution of the collision complex 
\cite{Zewail}.
Also, in  quantum kinetic theory of gases, accurate treatments must
abandon the ``completed collision''
approximation and use a non-asymptotic description, 
e.g. in terms of M\"oller wave operators instead of $S$ matrices,  as
in the Waldmann-Snider equation and its generalizations for moderately 
dense gases \cite{Snider90}. 

The theory has to adapt to these new trends by 
paying more attention to the temporal description of
the collisions.
Even if we restrict ourselves to asymptotic aspects, the cross section 
does not contain the whole information available in a
scattering process, since it is only proportional to 
the {\it modulus} of the $S$-matrix elements.
Information on the phase is 
available from delay times with respect to free motion. 
In fact, the full collision and not just the
asymptotic regimes should be understood     
to control or modify the products. This has      
motivated  a recent trend of theoretical and experimental
work to investigate the details of the interaction region 
and transient phenomena.  

In this chapter we restrict ourselves to one dimensional 
scattering. 
Many physical systems can be described in  
one dimension: the application of the effective mass
approximation to layered semiconductor structures
leads to effective one dimensional systems \cite{Bastard88};  
some surface phenomena are described by 1D models \cite{BE83}; 
and chemical reactions can in certain conditions be modelled by effective 
one dimensional potentials \cite{Levine}.   
The simplicity of 1D models have made them  
valuable as pedagogical and research tools. They facilitate 
testing hypothesis, new ideas, approximation methods and theories 
without unnecessary and costly complications.
For the same reasons they are frequently used to examine 
fundamental questions of quantum mechanics.
In particular, the time quantities 
treated in this book, such as tunneling or arrival times, 
have in most cases been examined in one dimensional models.  
Many results for 1D are inspired by results previously 
obtained in 3D, although  
the direct translation is not always 
trivial or possible. 3D collisions with spherically
symmetric potentials 
are described, by decomposition into partial waves, on the half line,
whereas 1D collisions involve 
the full line and a doubly degenerate spectrum. 

The chapter is organized as follows: 
Section 2 provides a minimal overview of  
formal 1D scattering theory. The treatment is ``formal''
because no mathematically rigorous proofs are given. Instead,  
we summarize the operator structure of the theory and the results
needed  to define characteristic times later on.
For a more rigorous mathematical presentation see e.g. \cite{DT79}.
Sections 3, 4 and 5 are devoted, respectively, to the dwell time,
the delay time, and 
decay times (the exponential decay and its deviations).
Quantities related to the tunneling time conundrum are scattered 
in several parts of the book.
In particular, sections 4.1 and 4.6 discuss
the Hartman effect and negative delays, 
while section 6 discusses 
the role of the Buttiker-Landauer ``traversal time''
in the time dependence of evanescent waves.    
A detailed discussion of the arrival times is left for 
Chapter .....

\section{Scattering theory in 1D}
\subsection{Basic premises and notation}
Let $H=H_0+V$ be the Hamiltonian operator for a single particle
in one dimension, where  
\beq
H_0=\frac{p_{\rm op}^2}{2m}
\eeq
is the kinetic energy operator in terms of 
the momentum operator
$p_{\rm op}$,\footnote{The subscript ``op'', meaning ``operator'', 
is not used for all
quantum operators, but only when confussion is possible with ordinary
numbers or functions.}
and 
$V$ is a ``local'' potential operator 
with coordinate  representation 
\beq
\la x|V|x'\ra=\delta(x-x')V(x)\;. 
\eeq
$V(x)$ must vanish for large values of $|x|$  
so that the M\"oller operators, defined below,
exist. 
This may certainly be accomplished by finite
range potentials, but spatial decays with infinite tails 
are also possible. 
       
The plane waves $|p\ra$, with coordinate representation given by
\beq
\la x|p\ra=h^{-1/2}e^{ixp/\hbar}\;, 
\eeq
are improper eigenstates\footnote{i.e., not in the Hilbert space of
square integrable states.}
of $p_{\rm op}$ and $H_0$,    
normalized according to Dirac's delta function,   
\beq
\la p|p'\ra=\delta(p-p')\;.
\eeq
Closure relations (or resolutions of the unit operator $1_{\rm op}$)
may therefore be written in momentum or 
coordinate representation as 
\beq
1_{\rm op}=\int_{-\infty}^{\infty} dx\,|x\ra \la x|=
\int_{-\infty}^{\infty} dp\,|p\ra  \la p|\;.
\eeq
\subsection{Basic abstract and parameterized operators.}
The state vector of the particle at time $t$
is denoted as $|\psi(t)\ra$
or simply as  
$\psi(t)$. We shall only deal with potentials such that at
large times in the past and future certain states $\psi$, the
scattering states,  
tend (in a strong sense)
to freely moving  asymtotic states 
$\phi_{\rm in}$ and $\phi_{\rm out}$ respectively,
\beqa
\psi(t)&\to&\phi_{\rm in}(t),\;\; t\to -\infty\;,
\\  
\psi(t)&\to&\phi_{\rm out}(t),\;\; t\to\infty\;.
\eeqa
The central objects in scattering theory are  
the {\it abstract} M\"oller operators. They link the asymptotic states
with $\psi$,
\beqa
\psi(t)&=&\Omega_+\phi_{\rm in}(t)\;,
\\
\psi(t)&=&\Omega_-\phi_{\rm out}(t)\;. 
\eeqa
Another important operator is  
\beq
S=\Omega_-^{\dagger}\Omega_+\;,
\label{Sdef}
\eeq
which links the two asymptotes,
\beq
\phi_{\rm out}(t)=S\phi_{\rm in}(t)\,.
\eeq
It is also convenient to introduce the auxiliary
``transition'' operators  ${\cal T}_\pm$ 
as 
\beq
{\cal T}_\pm=V\Omega_{\pm}\;.
\label{Tdef}
\eeq
The explicit definition of the M\"oller operators  
is given by infinite-time
(strong) limits,
\beq
\label{Mol}
\Omega_\pm=\lim_{t\to\mp\infty}e^{iHt/\hbar}e^{-iH_0t/\hbar}\;.
\eeq
The domain of these operators is the Hilbert space of  square
integrable 
states, although it is very useful to consider an
extension that can be applied on plane waves and allows to work
in a momentum representation.
To this end 
let us first define the {\it parameterized} operators
\beqa
\Omega(z)&=&1+G_0(z){\cal T}(z)\;,
\\
{\cal T}(z)&=&V+V G(z) V\;,
\eeqa
where $z$ is a complex variable with dimensions of energy, and   
$G(z)=(z-H)^{-1}$ and $G_0(z)=(z-H_0)^{-1}$ 
are the {\it resolvents} of $H$ and $H_0$.  
${\cal T}(z)$, $V$, $G(z)$ and $G_0(z)$  are also related by  
\beqa
G(z)&=&G_0(z)+G_0(z){\cal T}(z)G_0(z)
\\
{\cal T}(z)G_0(z)&=&VG(z)\;.
\eeqa
We shall see that the matrix elements of the resolvents
in coordinate representation are 
singular on the real positive axis, where there is a branch cut,
and at poles on the negative real axis (bound states).
Further singularities may occur by analytical continuation
on the second energy sheet. 

Note that the
operators of scattering theory have {\it abstract} or 
{\it parameterized} versions \cite{Snider88}.
Confusion may arise if they are not properly  
distinguished. 
The relation between abstract and parameterized operators is
found by acting with (\ref{Mol}) on a square integrable state. 
The resulting infinite-time limits can be substituted by the 
following limits, see e.g. \cite{Taylor72},    
\beqa
\Omega_+&=&\lim_{\vep\to 0+}\vep\int_{-\infty}^0 dt\,e^{\vep t}
e^{iHt/\hbar}e^{-iH_0t/\hbar}
\\
\Omega_-&=&\lim_{\vep\to 0+}\vep\int_{0}^\infty dt\,e^{-\vep t}
e^{iHt/\hbar}e^{-iH_0t/\hbar}\;.
\eeqa
Integrating, and introducing a closure relation in momentum,
\beqa
\Omega_{\pm}&=&\intf dp\, \Omega(E_p\pm i0)|p\ra\la p|\;,
\\
{\cal T}_{\pm}&=&\intf dp\, {\cal T}(E_p\pm i0)|p\ra\la p|\;.
\eeqa
The action of these operators on plane waves is now well defined. 
In particular, 
the improper eigenvectors of $H$ are obtained by acting with
the parameterized M\"oller operators
on the plane waves,
\beq\label{lis}
|p^{\pm}\ra=\Omega(E_p\pm i0)|p\ra=
|p\ra+\frac{1}{E_p\pm i0-H_0}{\cal T}(E_p\pm i0)|p\ra\;,
\eeq
where $E_p=p^2/(2m)$ 
is the energy of the
plane wave and of the corresponding eigenstate of $H$.  
This is the Lippmann-Schwinger integral equation for the states 
$|p^{\pm}\ra$, which are composed by a ``free'' plane wave 
and a ``scattering'' wave.   
To evaluate the coordinate representation and the asymptotic 
behaviour of the states at large distances, the matrix elements of the
free-motion resolvent are required,
\begin{equation}\label{resolvente}
\langle x|\frac{1}{E_p\pm i0 
-H_0}|x'\rangle=\mp\frac{im}{\hbar|p|}
\,e^{\pm i|p||x-x'|/\hbar}\,.
\end{equation}
(\ref{resolvente}) is obtained by introducing a resolution of unity in
momentum representation and using contour integration in the complex
momentum plane. 
Note that the two ways of approaching the real axis in (\ref{resolvente}),
from below or from above,
imply different boundary conditions at large $|x|$ for the two states in 
(\ref{lis}):  
the scattering wave 
of $|p^+\ra$ is formed by outgoing plane waves moving off the
potential region,
whereas the scattering wave of $|p^-\ra$ involves 
incoming plane waves towards the potential region.     

Since the plane waves $|p\ra$ form 
a complete set,  the 
following resolutions of
the operators $\Omega$, ${\cal T}$ and $S$ can be introduced,
\beqa\label{Om}
\Omega_{\pm}&=&\intf dp\, |p^{\pm}\ra\la p|\;,
\\
{\cal T}_\pm&=&V\intf dp\,|p^\pm\ra\la p|\;,
\\
\label{Sop}
S&=&\intf dp \intf dp'\, |p'\ra \la p'^{-}|p^+\ra\la p|\;.
\eeqa
Strictly speaking, the operators in (\ref{Om}-\ref{Sop}) are
not identical to the ones in  
(\ref{Sdef},\ref{Tdef},\ref{Mol}) since the domain of the former includes
plane waves. However, when acting on 
Hilbert space states they are  
equivalent so that, to avoid a clumsy notation, the same symbols will 
be used. A momentum representation is 
therefore allowed for these operators, which in general
involves distributions (generalized functions such as Dirac's delta or 
Cauchy's principal part). 

For real potential functions $V(x)$ the norm is conserved
throughout the 
collision, $\la \phi_{\rm in}|\phi_{\rm in}\ra=\la \psi|\psi\ra
=\la \phi_{\rm out}|\phi_{\rm out}\ra$. This means that the 
M\"oller operators are isometric, i.e.,  
\beq\label{iso}
\Omega_\pm^\dagger \Omega_\pm=1_{\rm op}\;.
\eeq
As a consequence, 
\beq
\la p^\pm|p'^\pm\ra=\delta(p-p')\;.  
\eeq
In general the M\"oller operators are not unitary because the bound
states are not 
in their range. Contrast this to the the operator $S$: it 
conserves the norm too, but  
it is unitary because it maps the whole Hilbert space onto the whole
Hilbert space, 
\beq
SS^\dagger=S^\dagger S=1_{op}\;. 
\eeq
The {\it scattering states} $\psi$ with incoming and outgoing
asymptotes move far away from the potential so they are
orthogonal to the bound states $\{|\Phi_j\ra\}$ at large
(positive or negative) times.  
Since the overlap amplitude $\la\psi|\Phi_j\ra=0$ is 
independent of time, the space of
bound states ${\cal B}$ is orthogonal to the scattering states, namely to the
range of the M\"oller operators. 
We shall always assume that the ranges of the two M\"oller
operators are equal to the subspace of scattering states ${\cal R}$, 
and that the
whole Hilbert space is the direct sum of the subspaces spanned by scattering 
and bound states, ${\cal H}={\cal R}\oplus{\cal B}$.
This assumption is known as 
{\it asymptotic completeness},   
\beq
\Omega_\pm\Omega_\pm^\dagger
=1_{\rm op}-\Lambda=\intf dp\,|p^\pm\ra\la p^\pm|\;. 
\eeq
In this expression the ``unitary deficiency'' $\Lambda$  
is the projector onto the subspace of bound
states,
\beq
\Lambda=\sum_j |\Phi_j\ra\la\Phi_j|\;.
\eeq
Taking matrix elements in (\ref{Sop}), the momentum representation 
of $S$ is given by   
\beq\label{spp}
\la p|S|p'\ra=\delta(p-p')-2i\pi\delta(E_p-E_{p'})
\la p|{\cal T}(E_p+i0)|p'\ra\;.
\eeq
The collision conserves the energy, which is,
asymptotically, kinetic energy. 
That is why $S$ commutes with $H_0$ and its matrix elements 
are proportional to an energy delta function.  
It is quite useful to factor out this delta function to  
define an on-the-energy-shell ${\bf S}(E)$ matrix.
Using
\beq
\delta(p-p')=\frac{|p|}{m}\delta(E_p-E_p')\delta_{pp'}\;,
\eeq
where $\delta_{pp'}$ is the Kronecker delta,
\begin{equation}
\delta_{pp'}=\cases{
1 &if $\,\, p=p'$\cr
0&if $\,\, p\ne p'$,\cr}
\end{equation}
and 
defining the matrix elements of ${\bf S}$, $S_{\alpha\beta}$, by 
\beq
\la p|S|p'\ra= 
|p| m^{-1}\delta(E_p-E_p'){S}_{\rm{sign}(p)\rm{sign}(p')}(E_p)\;,
\eeq
one finds      
\beq\label{spp'}
{S}_{\rm{sign}(p)\rm{sign}(p')}(E_p)=\delta_{pp'}-\frac{2i\pi m}{|p|}
\la p|{\cal T}(E_p+i0)|p'\ra\,,\;\;\;\;\;|p|=|p'|\;.
\eeq 
The subscripts $\alpha,\beta=\pm$ in the matrix elements 
${S}_{\alpha\beta}$,
denote the two possible 
``channels'', which correspond to positive ($+$) or negative ($-$)
momentum. 
A difference between the one dimensional scattering on the full line
($-\infty<x<\infty$) 
and radial scattering on the half line, ($0<r<\infty$) is that in 
the former, the ${\bf S}$ matrix is a unitary $2\times 2$ matrix while
in the later is a complex number of unit modulus.

\subsection{Symmetries}
\subsubsection{Time reversal invariance.}
This symmetry holds for real potentials. It implies 
\beq\label{simtri}
{S}_{\alpha\beta}={S}_{-\beta-\alpha}\;.
\eeq
%
%
\subsubsection{Parity.}
Frequently the potentials are symmetrical with respect to its
central position.
In that case,  
\beq
{S}_{\alpha\beta}={S}_{-\alpha-\beta}\;.
\eeq
\subsection{Eigenstates of $H$\label{eh}}
The eigenstates of $H$ given by the 
Lippmann-Schwinger integral equations (\ref{lis})
behave asymptotically
as a combination of two plane waves with positive and negative momenta.
The factors multiplying these plane waves are 
the {\it reflection and transmission amplitudes} according to the
following
table for asymptotic, long-distance behavior
(assume for the time being that $p>0$)  
\begin{equation}\label{pl}
\frac{1}{h^{1/2}}\cases{\exp(ipx/\hbar)
+R^l(p)\exp(-ipx/\hbar),&if $\,\, x\sim -\infty$\cr
T^l(p)\exp(ipx/\hbar),&if $\,\, x\sim\infty$,\cr}
\end{equation}
\begin{equation}\label{pr}
\frac{1}{h^{1/2}}\cases{
T^r(p)\exp(-ipx/\hbar),&if $\,\, x\sim -\infty$\cr
\exp(-ipx/\hbar)+R^r(p)\exp(ipx/\hbar),&if $\,\, x\sim \infty$.\cr}
\end{equation}
For potentials of finite range that vanish outside $[a,b]$ 
these are in fact exact expressions for $x<a$ and $x>b$.

If $p>0$, the boundary conditions in (\ref{pl})
define the states 
$\langle x|p^+\rangle$  
corresponding to an {\it incoming} plane wave
from the left, $\langle x|p\rangle$,
while 
the boundary conditions in (\ref{pr})
define the states $\la x|(- p)^+\ra$ 
corresponding to an {\it incoming} plane wave
from the right,
$\langle x|-p\rangle$. $T(p)$ and $R(p)$, with superscripts $l$ or $r$
for right or left incidence, are
the transmission and reflection amplitudes.
A wave packet peaked around a given 
$|p^+\ra$   
would be dominated by the plane wave $|p\ra$ before the collision, 
whereas, after the collision,
there would be two packets, one reflected and one 
transmitted with probabilities $|R(p)|^2$ and $|T(p)|^2$, dominated
by 
$|-p\ra$ and $|p\ra$ respectively, see e.g.
\cite{Diu80}.       

For $p<0$, however, the states determined by (\ref{pl}) and (\ref{pr})
correspond, respectively, to 
$\langle x|p^-\rangle$, with {\it outgoing} plane
wave $\la x|p\ra$, and $\langle x|(-p)^-\rangle$, with 
{\it outgoing} plane wave $\langle x|-p\rangle$.
A wave packet formed around $|p^-\ra$ would be close 
to a plane wave $|p\ra$ only {\it after} the collision occurs.
To form this 
peculiar outgoing state, the 
incoming asymptote must combine waves incident
from both sides of the potential barrier. This may of course be 
difficult to implement in practice, but it does not preclude the 
usefulness of these states as basis functions, and in general for 
applications where some control or selection of
the products of the collision is required.      

The previous discussion should make clear that 
$T(p)$, for $p<0$,  is {\it not} a standard transmission amplitude,
because it is {\it not} the amplitude of the transmitted plane 
wave of the state  
$|p^+\ra$, $p<0$. 
However, it analytically continues the standard transmission
amplitude ($T(p)$ for $p>0$) onto the $p<0$ domain, so the  
term ``transmission amplitude'' will be used 
irrespective of
the sign of $p$, even though the physical meaning is different for 
the two possible signs. According to our notational convention, positive
arguments of the amplitudes correspond always to states $|p^+\rangle$,
while negative momentum arguments correspond to $|p^-\rangle$ states.

\subsection{Relation between scattering amplitudes 
and basic operators}
Comparing the asymptotic (large $|x|$) behaviour of the states
in (\ref{pl}) and (\ref{pr})
with the asymptotic behaviour in (\ref{lis}),
the amplitudes $R(p)$ and $T(p)$ can be related to on-the-energy-shell
elements of the transition matrix.
We shall work out one case in detail: 
the scattering part of $\la x|p^+\ra$ for $p>0$ and
$x\to\infty$ is  
\beqa
&&\intf dx' \la x|G_0(E_p+i0)|x'\ra\la x'|
{\cal T}(E_p+i0)|p\ra
\nonumber\\
&&\sim
-\frac{2\pi m i}{h} \frac{e^{ipx/\hbar}}{p}\intf e^{-ipx'/\hbar}
\la x'|{\cal T}(E_p+i0)|p\ra
\nonumber\\
&&=-\frac{2\pi m i}{p} \la x|p\ra \la p|{\cal T}_+| p\ra\;.
\eeqa
Adding the free wave, $h^{-1/2}e^{ipx/\hbar}$, and comparing 
with (\ref{pl}), there results 
$T^l(p)=1-2i\pi m \la p|{\cal T}_+| p\ra/p$ for $p>0$.    
The rest of the cases can be worked out similarly 
(Because of time reversal invariance,
$\la p|{\cal T}_\pm| p\ra=\la -p|{\cal T}_\pm| -p\ra$,
and $T^r(p)=T^l(p)$.
Therefore the superscript for the transmission 
amplitude will be dropped hereafter):     
\beqa
T(p)&=&1-\frac{2i\pi m}{p}\la p|{\cal T}_{{\rm sign}(p)}| p\ra\;, 
\nonumber\\
R^l(p)&=&-\frac{2mi\pi}{p}\la -p|{\cal T}_{{\rm sign}(p)}| p\ra\;,
\nonumber\\
R^r(p)&=&-\frac{2mi\pi}{p}\;,
\la p|{\cal T}_{{\rm sign}(p)}| -p\ra\;.
\label{tp}
\eeqa
Some useful
relations follow from (\ref{tp}),
\beqa\label{tt}
[T(-p)]^*&=&T(p),\qquad\,\,\,\, p\,\,\, {\rm real}\;.
\\
\label{rr}
R^{r,l}(-p)^*&=&R^{r,l}(p)\,,\qquad p\,\,\, {\rm real}\;.
\eeqa
From (\ref{spp'}) and (\ref{tp}), the {\bf S} matrix is 
given by
\beqa
{\bf S}(p)\equiv{\bf S}(E)=\left(
\begin{array}{cc}
T(p)  &  R^r(p)\\
R^l(p) &  T(p)
\end{array}
\right)\,,\;\;\;p>0.
\label{smat}
\eeqa
It is quite useful to consider $\bf S$ as a (matrix) function of $p$.
In simple applications we only use
${\bf{S}}(p)$ with $p>0$,\footnote{Keep in mind that
$p>0$ in the arguments of ${\bf S}$ or of the 
scattering amplitudes does not mean ``incidence from the right''. 
According to the sign convention described in \ref{eh}, it means that
the amplitudes correspond to states with outgoing scattering parts:  
$|p^+\ra$ for right incidence, and $|-p^+\ra$ for left incidence.}  
but in fact we may also 
define ${\bf{S}}(p)$ for $p<0$ or even for complex $p$ 
in terms of the analytical continuations of the amplitudes $T(p)$, $R^r(p)$,
and $R^l(p)$. This extension will be discussed in sec. \ref{como}.

\subsection{The diagonal ${\bf S_d}$ matrix}
The ${\bf S}$ matrix (\ref{smat}) has been obtained from the momentum 
representation of $S$ using plane waves incident
form one side, $|\pm p\ra$, but other 
on-shell matrices may be defined in terms of a different  basis
formed by combinations of $|\pm p\ra$.
Of particular interest is the set  
$|u_j\ra,\; j=0,1,$ that provides a diagonal matrix  
\beqa
{\bf S_d}(p)=\left(
\begin{array}{cc}
S_0(p)  &  0\\
0 &  S_1(p)
\end{array}
\right)\;.
\label{smatd}
\eeqa
Unitarity implies that $|S_j|=1$, so the matrix elements may 
be written in terms of real eigenphase shifts $\delta_j$,
$S_j=e^{2i \delta_j}$. 
The $|u_j\ra$ 
are not mixed by the collision; these incident    
states produce an outgoing combination equal to
the incident one, except
for a phase factor. 
The diagonal ${\bf S_d}$ matrix is most advantageous for  
parity invariant potentials,
since the linear combinations become  
simply even and odd wavefunctions,
\beqa
|u_0\ra&=&{2^{-1/2}}(|p\ra+|-p\ra)\;,
\\
|u_1\ra&=&{2^{-1/2}}(|p\ra-|-p\ra)\;.
\eeqa
From the asymptotic behaviour of 
$|u_j^+\ra=\Omega_+|u_j\ra$ and $|\pm p^+\ra$ we may 
relate reflection and transmission amplitudes for even potentials
to the 
eigenphase shifts,  
\beqa
R(p)&=&2^{-1}\left(e^{2i\delta_0}-e^{2i\delta_1}\right)\;,
\\
T(p)&=&2^{-1}\left(e^{2i\delta_0}+e^{2i\delta_1}\right)\;.
\label{Tdel}
\eeqa
(Eq. (\ref{Tdel}) is in fact valid for arbitrary 
potentials.)
The boundary conditions for the  states $|u^+_j\ra$ are  
\beqa
\lim_{x\to-\infty}\la x|u_0^+\ra&=&e^{i\delta_0}\left(\frac{2}{h}\right)^{1/2}
\cos(-px/\hbar+\delta_0)\;,
\nonumber\\
\lim_{x\to\infty}\la x|u_0^+\ra&=&e^{i\delta_0}\left(\frac{2}{h}\right)^{1/2}
\cos(px/\hbar+\delta_0)\;,
\nonumber\\
\lim_{x\to-\infty}\la x|u_1^+\ra&=&ie^{i\delta_1}\left(\frac{2}{h}\right)^{1/2}
\sin(px/\hbar-\delta_1)\;,
\nonumber\\
\lim_{x\to\infty}\la x|u_1^+\ra&=&ie^{i\delta_1}\left(\frac{2}{h}\right)^{1/2}
\sin(px/\hbar+\delta_1)
\label{outer}\;.
\eeqa
It will be convenient for later manipulations to
drop the constant complex phase
factors and define real eigenfunctions of $H$ as
\beqa 
\la x|\psi_0\ra&=&e^{-i\delta_0}\la x|u_0+\ra\;,
\nonumber\\
\la x|\psi_1\ra&=&-ie^{-i\delta_1}\la x|u_1+\ra\;. 
\label{reale}
\eeqa
\subsection{Complex momentum\label{como}}
The properties of $T(p)$ as a function of the complex momentum $p$  
are of importance for many applications \cite{DT79}.
Let the potential function 
$V(x)$ be such that
\beq
\intf dx\,|V(x)|(1+x^2)  <\infty\;.
\eeq
Then $T(p)$ is meromorphic in ${\rm Im}\, p>0$
with a finite number $n_b$ of simple
poles
$i\beta_1$, $i\beta_2,...,,i\beta_n$, $\beta_j>0$, on the imaginary axis. 
The numbers $-\beta_j^2/(2m)$ are the eigenvalues of $H$. 
Moreover, 
\beq\label{largep}
T(p)=1+O(1/p)\;\;\;{\rm as}\;\;\;|p|\to\infty\,,\;\;{\rm Im}\, p\ge 0\;,
\eeq
and there can only be a zero at the real axis, at  $p=0$,
\beq
|T(p)|>0\;\;\; {\rm Im}\, p\ge 0,\;p\ne 0\;.
\eeq
In the generic case $T(0)=0$, and 
\beq\label{smallp}
T(p)=\gamma p+o(p),\; \gamma\ne 0,\; {\rm as}\; p\to 0,\;\;
{\rm Im}\, p\ge 0\;.
\eeq
Since $T(p)$ is meromorphic and it does not have zeros in the upper plane, 
the integral 
\beq
\frac{1}{2\pi i}\int_{\cal A} dp\,\frac{d \ln T(p)}{dp}=-n_b 
\eeq
along the contour ${\cal A}$
consisting of $[-R,-\epsilon]$, $[\epsilon,R]$, 
a semicircle of radius $\epsilon$ around the origin, and a large 
semicircle of radius $R$ in the upper half-plane,  
provides, according to a theorem of complex plane integration, 
the number of zeros (none in this case)
minus the number of poles of $T(p)$ enclosed (the bound states). 
The integral may also be evaluated using 
(\ref{tt}), (\ref{largep}), and (\ref{smallp}); this gives  
$2i\Phi_T(R)-2i\Phi_T(\epsilon)-i\pi$, where $\Phi_T(p)$ is the phase of 
$T$,
\beq\label{phT}
T(p)=|T(p)|\exp(i\Phi_T)\;.
\eeq
Combining the two results,  
\beq\label{lete}
\Phi_T(0)-\Phi_T(\infty)=\pi(n_b-{1}/{2})\,,
\eeq
which is {\it Levinson's theorem} for the case $T(p=0)=0$.
Otherwise, there is 
no $-i\pi$ contribution from the small semicircle and the phase difference 
becomes just $\pi n_b$. 
The convention followed is that $\Phi_T(\infty)=0$,  
so the theorem establishes the value of $\Phi_T(0)$.

The possibility to continue analytically $T(p)$ to the lower half plane
will depend of 
the potential considered \cite{MS96}.
Here we shall assume that
the continuation can be
performed (this is the case for example for potentials of finite range)
and discuss the properties that these continuations must obey. 
From  ${\cal T}^\dagger(z)={\cal T}(z^*)$
and the relations (\ref{tp}) we find 
\beqa
(R^{r,l}(p))^*&=&R^{r,l}(-p^*)\;,
\\
(T(p))^*&=&T(-p^*)\;, 
\eeqa
so that if there is a pole of $T(p)$ in the fourth quadrant at 
$p_R-ip_I$ ($p_R,p_I >0$),
there must be also a pole in the third quadrant at $-p_R-ip_I$. 
For an isolated pole,   
and if $p_I$ is small, the phase of $T(p)$ along the positive real
line will increase rapidly by $\pi$.
From (\ref{Tdel}) we see that poles of $T(p)$ are generally poles of $S_0$
or of $S_1$. Since $|T(p)|=|\cos(\delta_0-\delta_1)|$,
if the resonance eigenphase shift also jumps by $\pi$, while the other one
remains approximately constant, 
the transmission probability along the real axis 
will pass across a maximum ($1$),  or a minimum ($0$),  or both,
depending on the 
initial phase difference of the two eigenphase shifts. 
The above simplified picture will be blurred if the resonances are
very close to each other, or the pole is far from the 
real line.

\subsection{Unitarity and its consequences}
The unitarity of the collision ${\bf S}$ matrix,
${\bf S}{\bf S}^\da={\bf S}^\da {\bf S}={\bf 1}$, reflects
the conservation of norm in the collision. It  
provides two relations:
from the diagonal elements,
\beq\label{un1}
|T(p)|^2+|R^{r,l}(p)|^2=1\;,
\eeq
and from non diagonal ones, 
\begin{equation}\label{zero}
T(p)[R^l(p)]^*+[T(p)]^*R^r(p)=0\,,\;\;\; p\,\, {\rm real}\;. 
\end{equation}
Eq. (\ref{zero}) leads to a relation for the phases,
\beq\label{pha}
2\Phi_T-\Phi_{R^r}-\Phi_{R^l}=(2n+1)\pi\,\;\;n=0,\pm1,\pm2,...\;,
\eeq
where, as in (\ref{phT}), 
\beq
R^{r,l}(p)=|R^{r,l}(p)|e^{i\Phi_R^{r,l}(p)}\;. 
\eeq
\section{A measure of the collision duration: The dwell time}
In classical mechanics the quantity 
\beq\label{dwelltcl}
\tau_D(a,b;t_1,t_2)_{classical}=\int_{t_1}^{t_2}dt\int_a^b dx\,
\varrho(x,t)\;, 
\eeq
where $\varrho(x,t)$ is the probability density of an ensemble of 
independent particles, 
is the average over the ensemble of the time that each particle
trajectory spends between $a$ and $b$ within the time window
$[t_1,t_2]$ \cite{MBS92}. 
In other words, this is an average ``dwell'' or ``sojourn'' time in the 
selected space-time region.\footnote{The concept of a ``dwell time'' for 
a finite space region in the stationary regime 
is due to Buttiker \cite{Buttiker83}. Previously, integrals of the 
form (\ref{dwelltcl}) had been used to define time delays
by comparing  the free motion to that with a scattering center 
and taking the limit of infinite volume,
see e.g. \cite{GW75}} 
Its formal quantum mechanical counterpart is 
\beq\label{dwellt}
\tau_D(a,b;t_1,t_2;\psi)=\int_{t_1}^{t_2}dt\int_a^b dx\,
|\psi(x,t)|^2\,.
\eeq
In principle the coordinates $a$, $b>a$ and the instants  $t_1$
and $t_2>t_1$ are arbitrary but most often $a$ and $b$ are chosen
so that $V(x)$ is zero or negligible for $x<a$ and $b>a$.  
Hereafter $t_1$ will be, by default,  
$-\infty$, or occasionally $0$,
an initial preparation time; and $t_2=\infty$.

In spite of the formal similarity of the classical and quantum expressions,
the 
interpretation of (\ref{dwellt}) as a ``mean time'' spent in the 
region $[a,b]$, $[t_1,t_2]$ by quantum particles 
is not straightforward,  
since in the standard interpretation
of the quantum mechanical formalism there are no trajectories and 
therefore there is no obvious way to assign a time (duration) of
presence to a given member of the ensemble of particles associated 
with the quantum state.  
There are however several arguments 
that provide  (\ref{dwellt}) by extending to the quantum case the
classical dwell time, e.g. via Feynmann path integrals \cite{SC92},
causal or Bohm trajectories \cite{LA93}, or as an expectation value of a
hermitian sojourn time operator \cite{JW89}, see also Chapter ... for an
interpretation in terms of weak measurements.  
Irrespective of a hypothetical  statistical interpretation
of the dwell time in terms of individual members of
the ensemble, the dwell time
is at the very least a characteristic quantity of the ensemble represented 
by the state $\psi$, that quantifies the duration of the 
wave packet collision. For example, the dwell time
is considered an important parameter in high speed
applications of mesoscopic semiconductor
structures \cite{MT95}.

$\tau_D$ can be written in several ways, in particular 
as
\beq
\tau_D=\tau_D(a,b;-\infty,\infty)=\intf dt\,P_{ab}(t)
=\la \psi(t=0)|T_D|\psi(t=0)\ra\;,
\label{1}
\eeq
where  $P_{ab}(t)=\int_a^b dx\,\varrho(x,t)$,  
$T_D$ is the {\it sojourn time operator},  
\beq
T_D=\intf dt\, e^{iHt/\hbar}D(a,b)e^{-iHt/\hbar}\;,
\label{soj}
\eeq
and $D(a,b)$ is the projector onto the selected space region,
\beq
D(a,b)=\int_a^b dx\,|x\ra \la x|\;.
\eeq
An experimental determination of the  dwell time
may be carried out by monitoring the 
time evolution of the probability inside the selected spatial region, 
\cite{TMS87}.
This is admitedly an indirect route, where the first moment 
of $T_D$, $\tau_D$, is obtained without having measured individual 
dwell times for the members of the ensemble.      
It remains to be seen if second and higher moments of 
$T_D$ may be associated with some simple operational procedure. 

Let us now find other useful expressions for the dwell time.  
Integrating the continuity equation over $x$ between $a$ and $b$, 
and over time between $-\infty$ and $t$, $P_{ab}$ takes the form
\beq\label{5}
P_{ab}(t)=\int_{-\infty}^t dt'\,[J(a,t')-J(b,t')]
=\int_{-\infty}^t dt'\,\Delta J(a,b,t')\;, 
\eeq
where $J(x,t')$ is the current density, $\Delta J(a,b)=J(a)-J(b)$,
and the boundary condition $P_{ab}(-\infty)=0$ has been assumed.
Substituting (\ref{5}) into (\ref{1}), one finds
\beqa
\tau_D&=&\intf dt \int_{-\infty}^t dt'\,\Delta J(t')=\intf dt\intf dt'\,{\cal
H}(t-t') f(t')
\label{71}
\\
&=&\lim_{t''\to\infty}\int_{-\infty}^{t''} dt'\,(t''-t')\,\Delta J(t')
=\lim_{t''\to\infty}
\left[t''P_{ab}(t'')-\int_{-\infty}^{t''} dt'\,t'\,\Delta J(t')\right]\;.
\nonumber
\eeqa
Unless $P_{ab}(t)$
decays faster than ${t}^{-1}$, 
the dwell time will diverge.
The existence of a potential function leads generically  
to an asymptotic decay $\sim {t}^{-3}$, as discussed in 
section \ref{ltb}. However, for free motion 
the dwell time will diverge unless the momentum wave function 
vanishes at $p=0$, because of the dependence 
$\sim t^{1/2}$ of the free motion propagator, see 
(\ref{fmp}) below and the related discussion.   
In terms of the sojourn time operator (\ref{soj}) for $H_0$,
the possible divergence is due to a $|p|^{-1}$ factor, 
\beq
T_{D,H_0}=\sum_{\alpha=\pm}\int_{-\infty}^\infty dp\, \frac{mh}{|p|} 
|p\ra\la p|D|\alpha p\ra\la \alpha p|\;.
\eeq
In this and the following sections we shall limit ourselves in general to 
incoming asymptotes in the positive momentum channel ($+$)
that vanish 
at $p=0$, so  that the dwell time for free motion does exist. 
This will allow to compare dwell times with and without potential and   
to define delay times.     
These states, with a bounded support in momentum space, have necessarily 
a Fourier transform in coordinate space that can only 
vanish  at some set of points of measure zero. But this is not a
problem since 
the total probability for positive positions tends to zero as
$t\to -\infty$,  
\beq\label{prope}
\lim_{t\to-\infty}\int_a^\infty dx\,|\la x|\phi_{\rm in}(t)\ra|^2
=\int_{-\infty}^0 dp\,|\la p|\phi_{\rm in}(0)\ra|^2
\eeq
for any $a$ and any $\phi_{\rm in}$ \cite{JW87}.

Assuming that $tP_{ab}(t)\to 0$ as $t\to\infty$, the dwell time  
(\ref{1}) takes the local form 
\beq
\tau_D(a,b)=\intf dt'\,\left[J(b,t')-J(a,t')\right] t'\;.
\label{12}
\eeq
Other expression for states incident
in the positive momentum channel may be obtained by using resolutions
of the identity in terms of the states $|p^+\ra$,     
\beq
\tau_D(a,b;\psi)=\into dp\,|\la p|\phi_{\rm in}(0)\ra|^2\tau_D(p)\;,
\eeq
where
\beq
\tau_D(a,b;p)\equiv\frac{\int_a^b dx\,|\la x|p^+\ra|^2}{p/mh}\;,
\eeq
which suggests the interpretation of $\tau_D(a,b;p)$ as a dwell 
time for particles of definite momentum $p$ \cite{Buttiker83}.

Suppose now that $a<0$ and $b>0$
are both far from the barrier region,
before and after the barrier
respectively, so that 
the first passage of the wave packet across $a$   
can be described accurately in terms of the free motion asymptote 
$\phi_{\rm in}$, while the  
passage of the transmitted and reflected wave packets
can be evaluated with the  
asymptotic expressions   
\begin{eqnarray}\label{psit}
\psi_T(b,t)&=&{{1}\over{\sqrt{h}}}
\int_0^{\infty}\,{dp}\,\la p|\phi_{\rm in}(0)\ra\,T(p)
\,e^{i(pb-Et)/\hbar}\;,
\\
\label{psir}
\psi_R(a,t)&=&{{1}\over{\sqrt{h}}}
\int_0^{\infty}\,{dp}\,\la p|\phi_{\rm in}(0)\ra\,R(p)
\,e^{-i(pa+Et)/\hbar}\;.
\end{eqnarray}
(For a  potential with support between $0$ and $d$, $b$ could be taken
at the very barrier edge, $b=d$, but $a$ cannot be $0$, because    
of the strong interference between the incident and reflected parts.
$|a|$ should be much greater than the incident wave packet width in order to 
distinguish clearly the entrance passage from the reflected one.) 
Then, 
\beqa\label{16}
\intf dt'\,J_T(b,t')&=&\into dp\,|T(p)|^2|\la p|\phi_{\rm in}(0)\ra|^2=P_T\;,
\nonumber\\
\intf dt'\,J_I(a,t')&=&\into dp\,|\la p|\phi_{\rm in}(0)\ra|^2=1\;,
\nonumber\\
\intf dt'\,J_R(a,t')
&=&-\into dp\,|R(p)|^2|\la p|\phi_{\rm in}(0)\ra|^2=-P_R\;,
\eeqa
where the subscripts $I$, $T$ and $R$ in $J_I$, $J_T$ and $J_R$ mean that  
$\phi_{\rm in}$, $\psi_T$  and $\psi_R$ have been used to calculate
the fluxes. 
One can then write (\ref{12}) as 
\beq
\tau_D=P_T\la t\ra_b^{\rm out}-\la t\ra_a^{\rm in}
+P_R\la t\ra_a^{\rm out}\;,
\label{dwt}
\eeq
where
\beqa
\la t\ra_b^{\rm out}&\equiv&
\frac{\intf dt'\, J_T(b,t')\,t'}{\intf dt'\, J_T(b,t')}\;,
\label{18a}
\\
\la t\ra_a^{\rm in}&\equiv&\intf dt'\,J_I(a,t')\,t'\;,
\label{18b}
\\
\la t\ra_a^{\rm out}&\equiv&\frac{-\intf dt'\,
J_R(a,t')\,t'}{\intf dt'\,|J_R(a,t')|}\;.
\label{18c}
\eeqa
In each case the ``average passage instant''
is obtained by properly normalizing
the fluxes.
One may rightly wonder whether the notation and terminology used (as average
passage times) are justified.    
The ``averages'' are taken over the current density $J$,
a quantity that is not definite positive 
even for an incident wave packet without negative 
momentum components \cite{Allcock69,BM94,MPL99}.
It turns out, however, that the above ``averages'' 
over $J$ are equal to averages over a positively defined arrival 
time distribution (Kijowski's arrival time distribution) \cite{ML00},
as will be discussed in Chapter ... 
Models of detectors based on complex non-hermitian potentials lead 
also to these average times, delayed only by the small
(dwell) time that the particle spends in the detector before 
being detected \cite{MBM95}.
In the next section we shall relate these times to the ``phase times''.

Finally, note that (\ref{dwt}) could be, and has been, used to partition the 
dwell time into transmission and reflection components \cite{MBS92,BSM94}, 
see also the closely related aspproach of Olkhovsky and Recami \cite{OR92}. 
The main drawback is that the defined entrance average instant is common for 
both contributions, see \cite{DBM95,PMBJ97},
which is not a correct in the classical ensemble limit, 
and may lead to negative transmission 
times \cite{Leavens93} even in the classical case \cite{DBM95}. 
A two-detector model avoided this problem 
by assigning different entrance instants for
each member of the ensemble \cite{PMBJ97}.  
The distinction
between the dwell time and its components was first done by Buttiker   
\cite{Buttiker83}, and raised some controversy.  
As summarized in chapter 1, 
Muga, Brouard and Sala have  emphasized the multiplicity of possible quantum
partitionings versus the uniqueness of the  classical case, and 
developed a systematic theory to generate partitionings with 
the correct classical limit. Some of these  
include interference terms that cannot be assigned to transmission
or reflection but to both of them \cite{MBS92}.     
\section{Importance of the phases. Time delays.}
If the ${\bf S}$ matrix is known or simply one
of the amplitudes $R^l$ or $R^r$ is given as a function of momentum 
and there are no bound states, necessary and sufficient 
conditions are known for a unique potential to exist, and there are 
well established construction procedures \cite{Newton80,DT79}.
However a knowledge of the 
probabilities is not enough to determine the coefficients. 
The phases are associated with observable time dependent properties.

Consider a wave-packet impinging from the left on a barrier potential
located near $x=0$ (The exact barrier position is not important for our 
present purposes. Two typical choices for $x=0$ are 
the center of a symetrical barrier, or the left edge of a 
finite range potential). Let us   
take as before the spatial interval $[a,b]$ well outside
the barrier, so that there is a clear
separation between 
incoming and reflected passages.  
 
Since the incoming state is in the positive momentum channel,
\beq\label{phiin}
\la x|\phi_{\rm in}(t)\ra
=\into dp\, \la x|p\ra \la p|\phi_{\rm in}(0)\ra e^{(ipx-Et)/\hbar}\;,
\eeq
and, applying the M\"oller operator $\Omega_+$,
\beq
\la x|\psi(t)\ra
=\into dp\, \la x|p^+\ra \la p|\phi_{\rm in}(0)\ra e^{(ipx-Et)/\hbar}\;.
\eeq
(If the zero of time  
is taken well before the
wave packet interacts 
significantly with the barrier, the 
substitution $\la p|\psi(0)\ra=\la p|\phi_{\rm in}(0)\ra$
does not introduce any significant error.) 

Substituting (\ref{phiin}), (\ref{psit}) and (\ref{psir})
in the time averages (\ref{18a}-\ref{18c}),
and using the standard expression for the
current density, 
\beq
J(x,t)=\frac{\hbar}{m}\,{\rm Im}\,\!\!\left(\psi(x,y)^*
\frac{\partial\psi(x,t)}{\partial x}\right)\;,
\eeq
%
the derivative of an energy Dirac's delta may be identified 
and then used to perform one of the 
momentum integrals. 
The results are  
\begin{eqnarray}
\label{tbout}
\langle t\rangle_b^{out}&=&{{1}\over{P_T}}\int_0^{\infty}{dp}
\,|\la p|\phi_{\rm in}(0)\ra|^2\,|T(p)|^2\,{{m}\over{p}}\,\left[b
-x_0+\hbar\Phi'_T(p)\right]\;,
\\
\label{taout}
\langle t\rangle_a^{out}&=&{{1}\over{P_R}}\int_0^{\infty}{dp}
\,|\la p|\phi_{\rm in}(0)\ra|^2\,|R(p)|^2\,{{m}\over{p}}\,\left[-a
-x_0+\hbar\Phi'_R(p)
\right]\;,
\\
\label{tain}
\langle t\rangle_a^{\rm in}&=&\int_0^{\infty}{dp}
\,|\la p|\phi_{\rm in}(0)\ra|^2\,\frac{m}{p}[a-x_0]\;,
\end{eqnarray}
where the prime means derivative with respect to $p$, and 
\beq
x_0\equiv\hbar\,{\rm Im}\left(\la \phi_{\rm in}(0)|p\ra' 
\la p|\phi_{\rm in}(0)\ra\right)\;.
\eeq 
These results do not require to assume a narrow packet in
momentum representation.

The quantity 
\beq\label{taut}
\tau_T^{Ph}(x_0,b;p)\equiv m\left[b-x_0+\hbar\Phi_T'(p)\right]/p
\eeq
in the integrand of (\ref{tbout})  
consists of a time that a classical free particle with mass
$m$ and momentum $p$ would spend from $x_0$ to $b$, plus the
{\it time delay} $m\hbar\Phi_T'(p)/p$.
Similarly, the term in brackets in (\ref{taout}),
\beq\label{taur}
\tau_R^{Ph}(x_0,a;p)\equiv m\left[-a-x_0+\hbar\Phi_R'(p)\right]/p\;,
\eeq
is a
time spent by a classical particle that travels freely from $x_0$ to $x=0$,
where its momentum is
instantly reversed, and from $x=0$ to $a$, plus a delay contribution. 
It is to be noted that unless $a=-b$ the reference time associated with 
classical free motion is 
different in the transmission and reflection cases. We    
shall see a consequence of this disparity afterwards
when calculating  average delays in \ref{4.2}.
  
{\it Formally} we may use (\ref{taut}) and (\ref{taur}) to 
{\it define} ``phase times'' for arbitrary values of 
$a$, $b$, and $x_0$.   
In particular, for   
a finite range barrier between $x=0$ and $d$  
let us define
\beq
\tau^{Ph}_T(0,d;p)=\frac{md}{p}+\frac{m\hbar}{p}\Phi_T'(p)\;,
\eeq
by substracting from $\tau^{Ph}_T(x_0,d;p)$
the classical flight time
between  $x_0$ and $0$, $-mx_0/p$. These 
``extrapolated phase times'' for traversal
should not be overinterpreted
as actual traversal times \cite{LA89,HS89}, not only because, as 
pointed out in Chapter 1, there is not a unique traversal time, but because  
a wave packet peaked around $p$ is very broad in coordinate
representation, so it is
severely deformed before the hypothetical
``entrance'' instant $t_{ent}=|x_0|m/p$, and at $x=0$ there is an important
interference effect between incident and reflected components.   
The wavefunctions 
$\phi_{\rm in}$ and $\psi_R$ used to calculate the fluxes 
$J_I$ and $J_R$ do not faithfully represent the
actual wave, so that the average instants (\ref{taout},\ref{tain})
loose their physical meaning as average detection times.   

\subsection{The Hartman effect}
Relation (\ref{tbout}) is suitable for examining
the ``Hartman effect'' \cite{Hartman62,Fletcher,LA89,OR92,BSM94}.
Hartman \cite{Hartman62} studied the
evolution of a wave-packet with momentum distribution centered
around $p_c$, colliding with a rectangular barrier of height
$V_0>p_c^2/(2m)$, and width $d$.
He found three regions according to the value of $d$.
For large barrier widths 
(opaque barrier conditions), the
stationary phase time associated with $p$, under the barrier, goes
to a constant,
$\tau^{Ph}_T(x_0,d;p)=2m/(p\kappa)-x_0m/p$, independent of $d$, where 
\beq
\kappa=[2m(V_0-E)]^{1/2}/\hbar\;.
\eeq
When transmission is dominated by momentum
components below the barrier, the transmitted wave-packet seems
to traverse the potential
region in a time interval independent of $d$.
This is the ``Hartman effect''.
If $d$ is increased further,  
plane waves with momentum above the barrier height dominate the
transmission, and classical behaviour results, i.e., time
grows linearly with $d$.
Finally, for small barrier widths, Hartman defined a ``thin barrier
region'' where the phase time depends generally on $d$.

To be more specific,  
let us consider the initial Gaussian wave-packet
\begin{equation}\label{psi0}
\la x|\phi_{\rm in}(0)\ra=\left[{{1}\over{2\pi \delta^2}}\right]^{1/4}
\exp{\left[ip_c x/\hbar
-(x-x_c)^2/(4 \delta^2)\right]}\;,
\end{equation}
of average momentum $p_c=\hbar k_c$ and spatial width
(square root of the variance) 
$\delta$. Here  
$x_0$ becomes equal the wave packet center $x_c$.
The initial momentum
distribution is a Gaussian distribution with 
variance $\sigma^2=[\hbar/(2\delta)]^2$.
We assume that $p_c>>\sigma^2$ so that the truncation 
at $p=0$ in (\ref{tbout}) is not significant.  
For an energy
distribution peaked around $E_c<V_0$ the following results can
be drawn \cite{BSM94}:

If $\kappa_c d\equiv\sqrt{2m(V_0-E_c)}d/\hbar>>1$, 
$\langle t\rangle_d^{out}$ does not
vary appreciably when $d$ increases, thus showing Hartman effect.
When $d$ is sufficiently large, the components of the wave-packet
under the barrier are so strongly depressed by $|T(p)|^2$ that higher
momenta start to dominate, and $\langle t\rangle_d^{out}$ grows
almost linearly, as one expects classically. 
As $\delta$ is increased, larger values of $d$ are needed to
pass from the first regime to the second one. An estimation
of the value of $d$ which gives the transition between
Hartman effect and quasiclassical behaviour can be obtained for each
value of $\delta$ by equating the factor
$|T(p)|^2\,|\la|\phi_{\rm in}(p)\ra|^2$
for $p=p_c$ and for $p=p_r$, where $p_r$
is the momentum of the first resonance above the barrier. 
This  leads to the relation
\begin{equation}\label{dim}
\delta={{\hbar\sqrt{-\ln{|T(p_c)|}}}\over{|p_r-p_c|}}\approx
{\hbar{\sqrt{\kappa_cd}}\over{|p_r-p_c|}},
\end{equation}
between $\delta$ and $d$, that 
clearly separates quantum and quasiclassical behaviour.
Also, for fixed $\delta$, the transition
is sharper at larger $\delta$ as a consequence of the 
narrower momentum distribution.  

We have already warned the reader against a naive overinterpretation of the
extrapolated phase time $\tau^{Ph}_T(0,d;p)$,
which becomes $2m/(p\kappa)$ for the barrier
traversal in the Hartman effect, mainly because of the strong deformation
of the broad incident wavepacket. We could try to avoid the interpretational
pitfalls of this quantity and look instead at the time $\la t\ra_d^{out}$
for a wave packet which is initially localized near the edge of the 
barrier, and with a small spatial width compared to the barrier length $d$,
to identify the entrance time and the preparation
instant with a tolerable small uncertainty. However, Low and Mende
speculated \cite{LM91}, and then Delgado and Muga 
have shown \cite{DM96}, that this localization leads to the dominance of
over-the-barrier components.
Similar conclusions are drawn from a two detector model, one before and
one after the barrier, when the detector before the barrier  localizes the 
particle into a small
spatial width
compared to $d$ \cite{PMBJ97}.

\subsection{The lifetime and delay time matrices\label{4.2}} 

The four delay times corresponding to reflection and transmission for 
right and left incidence form the 
{\it delay time matrix}
introduced by Eisenbud in his thesis \cite{Eisenbud48},  
\beq
\Delta t_{\alpha\beta}=
{\rm{Re}}\left[-i\hbar\frac{1}{S_{\alpha\beta}}
\frac{dS_{\alpha\beta}}{dE}\right]\;.
\eeq
The matrix element $\Delta t_{\alpha\beta}$ is the delay time in the 
appearance of the peak outgoing signal in channel $\beta$, after 
the injection of a pulse narrowly peaked in momentum in channel $\alpha$.  
The ``delay'' may in fact become negative as discussed already.      
These delay times have been traditionally
obtained by means of the ``stationary phase
argument''.  Let us rewrite 
the transmitted wave function as  
\beq
\la x|\psi_T(t)\ra = h^{-1/2}\into dp\, e^{ixp/\hbar-iE_pt/\hbar+i\Phi_T}
\la p|\phi_{\rm in}(0)\ra |T(p)|\;.    
\eeq
If the initial state is narrowly peaked around
$p_0$, the integral will be appreciably different from zero only if the phase
of the exponential function is stationary near $p=p_0$. This implies a 
``spatial delay'' with respect to the free-motion wave packet, 
\beq
\Delta x= \hbar\frac{d\Phi_T}{dp}\bigg|_{p=p_0}\;,
\eeq
and a corresponding ``time delay''
\beq
\Delta t_{++}(p_0)=\frac{\hbar m}{p_0}\frac{d\Phi_T}{dp}\bigg|_{p=p_0}\;.
\eeq
The time delays are also related to the  
on-the-energy-shell lifetime matrix of Smith \cite{Smith60}, 
\beq\label{Q}
{\bf Q}(E)=i\hbar{\bf S}(E)\frac{d{\bf S}(E)^\da}{dE}\;,
\eeq
${\bf S}$ is unitary, so ${\bf Q}$ is Hermitian. 
Thus the diagonal matrix elements of ${\bf Q}$ are real 
and take the form 
\beq
{Q}_{\alpha\alpha}=\sum_\beta |{S}_{\alpha\beta}|^2
\Delta t_{\alpha\beta}\;.
\eeq
Since the particle has a probability $|{S}_{\alpha\beta}|^2$
to emerge in the 
channel $\beta$, ${Q}_{\alpha\alpha}$ is the average
delay experienced by 
the particle injected in channel $\alpha$.

We shall now relate the ${\bf Q}$ matrix with the ``wave packet 
lifetime'', defined as the difference between dwell times with 
and without potential \cite{Smith60,Sassoli93}, 
\beq
\la Q\ra\equiv\tau_{D,\psi}
-\tau_{D,\phi_{\rm in}}\;.
\eeq
As before, the incidence is in the positive momentum channel.  
$\tau_{D,\psi}$ is given by (\ref{dwt}) whereas 
the dwell time for free motion is 
\beq
\tau_{D,\phi_{\rm in}}=\la t\ra_{b,\phi_{\rm in}}^{out}
-\la t\ra_{a,\phi_{\rm in}}^{in}
=\int_0^{\infty}{dp}\,|\la p|\phi_{\rm in}(0)\ra|^2\,\frac{m}{p}[b-a]\;,
\label{tauphi}
\eeq
where, similarly  to (\ref{tain}), 
\begin{equation}\label{tbout'}
\langle t\rangle_{b,\phi_{\rm in}}^{\rm out}=\int_0^{\infty}{dp}
\,|\la p|\phi_{\rm in}(0)\ra|^2\,\frac{m}{p}[b-x_0]\;.
\end{equation} 
Since, by hypothesis,   $\la t\ra_{a,\psi}^{in}
=\la t\ra_{a,\phi_{\rm in}}^{in}$, $\la Q\ra$ takes the form
\beqa
\la Q\ra &=&\into dt\, \int_{a}^b dx 
\left(|\la x|\psi(t)\ra|^2-|\la x|\phi_{\rm in}(t)\ra|^2\right)
\nonumber\\
&=&P_T[\la t\ra_{b,\psi}^{out}-\la t\ra_{b,\phi_{\rm in}}^{out}]
+P_R[\la t\ra_{a,\psi}^{out}-\la t\ra_{b,\phi_{\rm in}}^{out}]\;.
\eeqa
Substituting all the integral expressions
obtained for the passage times,
and writting $c=-a-b$, 
\beq
\la Q\ra= \hbar\int_0^\infty dp\, \frac{m}{p}
|\la p|\phi_{\rm in}(0)\ra|^2\left[\Phi_T'|T(p)|^2+
\left(\Phi_R'+\frac{c}{\hbar}\right)
|R(p)|^2\right]\;.
\eeq
Note the term proportional to $c$ in the reflection 
part. It arises because of the mismatch between 
the free motion reference times used to define the 
reflection and transmission time delays when $c\ne 0$. 
Choosing $c=0$,     
$\la Q\ra$ represents the weighted momentum average of the 
mean delay for each momentum,\footnote{Additional oscillatory
terms, see e.g. \cite{HS89,Nussenzveig2000}, appear
when the no-interference condition between the reflected
and incident wave packets
is not imposed.}  
\beq\label{avq}
\la Q\ra=\into dp\,|\la p|\phi_{\rm in}(0)\ra|^2 {Q(E)}_{++}\;.
\eeq
The eigenvalues of $\bf Q$ have been used as good indicators of 
resonances \cite{KK81}, see \ref{BW} below, 
and may be interpreted for symmetrical potentials  as 
the delays associated with symmetrical or antisymmetrical bilateral
incidence 
\cite{Nussenzveig2000}.  
However their operational interpretation in terms of individual measurements
is puzzling.   
An asymptotic measurement of the arrival time at $b$
in the transmission side could be done
in principle for one of the
identically prepared systems
represented by the wave packet.
Because of the coordinate spread of the wavepacket, however, 
there is a large uncertainty in the 
time that the {\it same}
particle enters the region $[a,b]$. If  
a detector is placed at $a$ before
the collision occurs, the entrance  time can be determined, but  
in general either the particle is destroyed or its behaviour afterwards
is modified by the measurement. 
We are thus faced with an intrinsic difficulty to measure
{\em individual} delays. 
This means that, at variance with
other quantum mechanical averages which are
interpreted  as averages of the eigenvalues  measured for the individual 
members of the ensemble, the operational meaning of (\ref{avq})  
does not require to assign a lifetime 
to a given particle. It depends on   
the average times defined in (\ref{18a}-\ref{18c}), which are 
measurable, 
at least in principle, by the time-of-flight technique  
(Other operational procedure making use of particle absorption
along the chosen interval has been described by Golub et al.
\cite{GFGG90}). 
This peculiarity of the delay time was
already noted by Goldrich and Wigner \cite{GW72}.  
A consequence is that the ordinary quantum fluctuations around the 
average value are not operationally meaningful. Instead, 
the relevant fluctuations refer to variations of the average
values themselves, 
corresponding to ${\bf S}$ matrix (or Hamiltonian) ensembles \cite{MW95}. 

The trace of (\ref{Q}) in the on-shell space
is related to the change in density of states 
$\Delta \rho(E)\equiv\tr [\delta(E-H)
-\delta(E-H_0)]$  which is a fundamental quantity to characterize the 
continuous spectrum \cite{dest} according to the  
``spectral theorem'' (The three dimensional elastic and multichannel
versions of the spectral
theorem have been extensively discussed and proven rigorously
\cite{spect}.)
\beqa
\Delta \rho(E)&=&-\pi^{-1}
{\rm Im}\, \tr [G(E+i0)-G_0(E+i0)]\nonumber
\\
&=&\frac{1}{h}\sum_{\alpha}{Q}(E)_{\alpha}=
\pi^{-1}\frac{d\Phi_T(E)}{dE}\,.
\label{sth}
\eeqa
The second equality (spectral theorem) 
follows from a result of Dashen, Ma and Bernstein
\cite{DMB69}. 
To obtain the final expression, use has been made of (\ref{un1}) and
(\ref{pha}) \cite{AB85}, see \cite{TG93} for an alternative derivation 
consisting in evaluating $\Delta \rho$
for a finite system and then going to 
infinity. 
Note that the maxima of the trace of $\bf{Q}$ may be used to identify
resonance energies and widths \cite{JCL94}.    
For further relations between the density of states and the dwell time
see \cite{GP93,Iannaccone95,IP96}.  
Chapter ... discusses the concept of local density of states 
and its relation to the Larmor clock and transport properties.  

\subsection{Breit-Wigner resonances\label{BW}}
The simplest model of resonance behaviour
is the Breit-Wigner model for an isolated 
resonance,
\beq
{\bf S}(E)=1-\frac{i{\bf A}}{E-E_0+i\Gamma/2}\;.
\eeq
By imposing unitarity to ${\bf S}$, and assumming that ${\bf A}$ and 
the resonance parameters 
$E_0$ and $\Gamma$ are independent of $E$,
it follows that 
${\bf A}={\bf A}^\dagger$ and  
\beq\label{a2g}
{\bf A}^2=\Gamma {\bf A}\;.
\eeq
This means that the matrix $\bf{A}$ factorizes as 
$A_{\alpha\beta}=\gamma_\alpha\gamma_\beta^*$, and that  
it is proportional to a projector matrix ${\bf P}={\bf A}/\Gamma$ with 
eigenvalues $1$ and $0$.
Thus, the equation (\ref{a2g}) takes the form
\beq
\Gamma=\sum_\alpha|\gamma_\alpha|^2\;. 
\eeq
The corresponding ${\bf Q}$ matrix may now be written as 
\beq
{\bf Q}={\bf P} q_m\;,
\eeq
with eigenvalues $q_m$ and zero, where  
\beq
q_m=\frac{\hbar \Gamma}{(E-E_0)^2+\Gamma^2/4}
\eeq
is the maximum value allowed for a diagonal element of ${\bf Q}$.  
The Breit-Wigner model for ${\bf S}$ and ${\bf Q}$
can be generalized in various ways, in particular to account 
for multiple overlapping resonances \cite{MW95}.

\subsection{Negative delays}

\begin{figure}[tbp]
\begin{center}
\includegraphics[width=.6\textwidth]{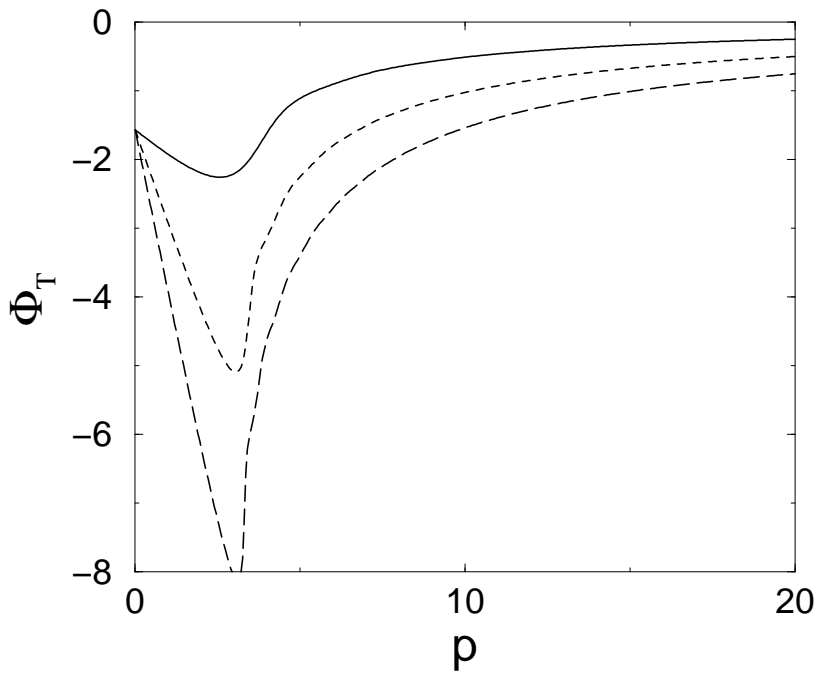}
\end{center}
\caption[]{Phase of the transmission amplitude versus momentum for a square 
barrier of ``height'' $V_0=5$ and for three 
different widths, $d=1$ ({\it solid line}), $2$ ({\it short dashed line}),
and $3$
({\it long dashed line}). $m=1$. (all quantities in atomic units)} 
\label{tdelayph}
\end{figure}

In partial wave analysis of three dimensional collisions 
with spherical potentials, the time delay has been used mainly
as a way to characterize 
resonance scattering. One of the standard definitions of a 
resonance is a jump by  $\pi$ in the eigenphases of the
$\bf S$ matrix. 
In one dimensional collisions 
the time delay has been also used frequently to characterize      
(non-resonant) tunnelling, where it may become negative. In fact the 
different delay signs associated with the two 
types of effects, resonances and tunnelling, are not independent.      
In 3D it was soon understood by Wigner \cite{Wigner55} that the
increases and 
decreases of the phase should balance each other. Since Levinson's 
theorem imposes a fixed phase difference from $p=0$ to $\infty$, 
there must be intervals of negative delay 
to compensate for the phase increases associated
with the resonances. 
A similar analysis applies in 1D to the transmission amplitude. 
In Figure \ref{tdelayph},
the phase of the transmission amplitude for a square 
barrier is shown 
versus $p$ for different values of the barrier width $d$. 

As $d$ increases, the scatering resonances ``above the barrier''
$p>p_0=(2mV_0)^{1/2}$ become more dense and are defined better because 
of the approach of the resonance poles in the fourth quadrant to the 
real axis. The corresponding
increases of the phase are compensated by a more and more negative  
delay in the tunneling region.  

Negative delays also arise if a pole of $T(p)$ crosses the real axis
upwards, when varying the interaction strength, 
to become a loosely bound state in the positive imaginary axis.
Levinson's theorem, see (\ref{lete}), imposes then a sudden jump in the phase 
$\Phi_T(0)$ that must be compensated by a strong negative slope.
This effect is more important near threshold, i.e., when the  
pole is very close to the real axis \cite{DK92}.
Similar effects have been described for non-bound state poles in 
complex potential scattering \cite{MP98}.

Wigner also found a bound for the negative  
(partial wave) delay time of a potential 
of finite radious.
Whereas positive delays can be arbitrarily large, negative
delays are restricted by ``causality conditions'' \cite{Nussenzveig72}.
Some back-of-the-envelope causality arguments may however be  
misleading. For example, assume a barrier of length $d$, 
and let $a$ coincide with the left edge and $b$ with the right 
edge. If the total time $\tau_T^{Ph}(0,d)$ is to be
positive, the delay cannot be more negative than the reference free
time, 
\beq\label{bound}
\Delta t_{++}>-\frac{md}{p}\;,
\eeq
see e.g. \cite{GP90}. 
In fact this bound may be violated, in particular at low energy 
in the proximity of 
a loosely bound state. This should not surprise the reader after our 
repeated warnings against an overinterpretation of the extrapolated time 
$\tau^{Ph}_T(0,d)$. The flaw in the argument is the assumption of 
positivity of $\tau_T^{Ph}$.  
Nevertheless, rigorous bounds have been established by Wigner himself and 
various authors in 3D collisions, 
see \cite{Martin81,Nussenzveig72} for review. 
In 1D collisions  the following bound holds  
for even potentials with finite support between 
$-b$ and $b$ \cite{DK92,Sassoli94}: 
\beqa
\Delta t_{++}&\ge&\frac{m}{p}   
\left\{-2b+\frac{1}{4p}[\sin(2pb/\hbar+2\delta_0)-
\sin(2pb/\hbar+2\delta_1)]\right\}
\nonumber\\
&\ge&
\frac{m}{p}\left(-d-\frac{1}{2p}\right)\;.
\label{DiKi}
\eeqa
This may be proven by using the even and odd eigenfunctions 
$\la x|\psi_{j}\ra$ introduced in (\ref{reale}), 
in particular the fact that 
$\int_{-b}^b dx\,\psi_j^2>0$. 
We start by calculating the logarithmic derivative of $\la x|\psi_0\ra$  
at $x=b$ from the known expression
for the outer region, see (\ref{outer}), 
\beq
L_b\equiv\frac{d\la x|\psi_0\ra/dx}{\la x|\psi_0\ra}\bigg|_{x=b}=
-\frac{p}{\hbar}\tan(pb/\hbar+\delta_0)\;.
\eeq
Taking the derivative of $L_b$ with respect to $p$, 
\beq
\frac{d\delta_0}{dp}=-\left\{\frac{\hbar}{p}\frac{dL_b}{dp}
\cos^2(pb/\hbar+\delta_0)
+\frac{1}{2p}\sin[2(pb/\hbar+\delta_0]+\frac{x}{\hbar}\right\}\;.
\eeq
The first term on the right hand side may also be written as
\beq
\frac{\hbar}{m}[\la x|\psi_0\ra_E\la x|\psi_0\ra_x
-\la x|\psi_0\ra\la x|\psi_0\ra_{E,x}](x=b)\;,
\eeq
where the subscripts $E$ and $x$ are shorthand notation for the 
derivatives with respect to $E$ and $x$.
Repeating the same operations for $x=-b$
one finds that  
\beqa
&&[\la x|\psi_0\ra_E\la x|\psi_0\ra_x-\la x|\psi_0\ra
\la x|\psi_0\ra_{E,x}](x=b)
\nonumber\\
&=&-[\la x|\psi_0\ra_E
\la x|\psi_0\ra_x-\la x|\psi_0\ra\la x|\psi_0\ra_{E,x}](x=-b)\;.
\label{b-b}
\eeqa
We shall now prove that this is a positive quantity.
Taking the derivative of the stationary Schr\"odinger equation
with respect to 
energy one obtains for real eigenfunctions of $H$
the identity \cite{Smith60}  
\beq
\la x|\psi\ra^2=-\frac{\hbar^2}{2m}\frac{\partial}{\partial x}
\left(\la x|\psi\ra \la x|\psi\ra_{E,x}
-\la x|\psi\ra_E \la x|\psi\ra_{x}\right)\;,
\eeq
so that, using (\ref{b-b}),  
\beq
\int_{-b}^{b}dx\,\la x|\psi_0\ra^2=\frac{\hbar^2}{m}
\left(\la x|\psi_0\ra_E\la x|\psi_0\ra_x-\la x|\psi_0\ra
\la x|\psi_0\ra_{E,x}\right)(x=b)\;.
\eeq 
Carrying out similar manipulations for the odd wavefunction 
$\la x|\psi_1\ra$, (\ref{DiKi}) is found as a consequence of the  
positivity of the probability to find the particle in
the barrier region. 

According to this bound the negative delay may be
arbitrarily large for small enough momenta and may diverge at 
$p=0$, as it occurs when a bound state appears when making the 
potential more atractive \cite{DK92}. For the square barrier, which does not
have bound states,  the time advancement 
of the Hartman effect is less important and it 
is actually bound by (\ref{bound}).
Thus, whereas the experiments looking for anomalously large 
traversal velocities (``superluminal effects'') have been frequently
based on evanescent conditions in square barriers (tunneling),  
square wells with the proper depth may in fact lead 
to much larger advancement effects at threshold energies.

\section{Time dependence of survival probability: Exponential decay
and deviations}

    The quantum mechanical decay of unstable states can be
described in many different ways \cite{WW,AJP}.
In many theoretical works the emphasis has
been
on justifying the approximately valid exponential decay law.
A possible treatment for the survival amplitude
$A(t,\psi)\equiv\langle\psi (0)|\psi (t)\rangle$ decomposes the state
$\psi$ by the usual resolution into proper and improper eigenstates of
the Hamiltonian $H$, corresponding to bound and
continuum states.
Even though it contains all the information, this is not convenient in
general either for calculational purposes or for rationalizing the
decay behaviour in a simple manner, except in favourable circumstances
where the integral is easily approximated and parameterized, e.g. for
isolated resonances and particular initial states. An ideal
description would handle arbitrarily complex initial states and
potentials {\it in simple terms}, and allow for an understanding of both
the dominant exponential decay and the deviations from it.

        Much progress in this direction has been achieved by
representating $A(t,\psi)$ as a discrete sum over resonant terms
\cite{GCMM95,MWS96}. The discretization allows a clear identification and
separation of the physically dominant contributions, different terms
being important for different time regimes. 
Here we follow the treatment presented in \cite{MWS96,MSW96}.      

   The survival amplitude $A(t,\psi)=\langle\psi(0)|\psi(t)\rangle$
requires the diagonal matrix elements of the unitary evolution
operator $e^{-iHt/\hbar}$. When this operator is expressed in terms of
the resolvent, $A(t,\psi)$ takes the form
\beqa
A(t,\psi)&=&\langle \psi|e^{-iHt/\hbar}|\psi\rangle
\nonumber\\
&=&
\label{S}
\frac{i}{2\pi m}\int_{\cal C}
dq\, q \langle\psi|\frac{e^{-izt/\hbar}}{z-H}
|\psi\rangle={i\over2\pi}\int_{\cal C}dq\, e^{-izt/\hbar}M(q)\;,
\eeqa
where $z=q^2/2m$ is a complex energy and the contour ${\cal C}$
goes from $-\infty$ to
$+\infty$  passing above all the singularities of the resolvent due to
the spectrum of $H$ (discrete poles for bound states and the natural
boundary of the real axis for the continuum), and
\begin{equation}\label{Iq}
M(q)\equiv\frac{q}{m}\langle\psi|\frac{1}{z-H}|\psi\rangle\;.
\end{equation}
The survival probability is to be calculated as
${\cal S}(t,\psi)=|A(t,\psi)|^2$.

\subsection{Predicted time behaviour}

The function $M(q)$
is evaluated in the upper half $q$-plane and then analytically
continued into the lower half plane. Provided that the continuation
exists, $M(q)$ has in general a set of {\it core} singularities,
depending only on the potential, plus possibly  other 
{\it structural}
state-dependent singularities. 
It is then useful to deform the original integration contour
to the diagonal ${\cal D}$ of the second and fourth quadrants of
the $q$-plane. This provides both physical insight by identifying the
most relevant time dependence (exponential decay) of the survival, and
a calculational advantage for the remainder, since for $t>0$ the
exponential $e^{-izt/\hbar}=e^{-iq^2t/(2m\hbar)}$ is a real Gaussian
on this diagonal.

        Let us assume that a pole expansion of the form
\begin{equation}
M(q)=\sum_{k} \frac{a_k}{(q-q_k)}
\end{equation}
is possible (higher order poles
can be treated in a similar fashion).  Here $k=1,2,3\cdots$
indexes the poles. On
deforming the $q$ integration from contour ${\cal C}$ to
${\cal D}$, the residues of the poles
$q_k$ crossed in the fourth quadrant on carrying out this
deformation provide contributions to $A(t)$ 
that decay exponentially with time, whereas 
the residues are 
purely oscillatory for poles in the upper half plane (bound states), 
\begin{equation}
E_k(t)=a_k e^{-iq_k^2 t/(2m\hbar)}=a_k e^{-u_k^2}\;,
\end{equation}
where
\beq\label{uf}
u\equiv q/f,\,\,\,\,\,f\equiv(1-i)\sqrt{(m\hbar/t)}\;.
\eeq
becomes real along the diagonal ${\cal D}$. 
Independently of providing or not providing a residue, 
all poles contribute because of the integral along the
diagonal. Each pole contribution is 
expressed in terms of the $w$-function, see \cite{AS72}, as 
\begin{equation}\label{d}
D_k(t)=-\frac{a_k}{2}{\rm sign}({\rm Im} u_k) \,w[{\rm sign}({\rm{Im}} u_k)
u_k]\;.
\end{equation}
The exponential term may be added to this contribution to
give the compact result \cite{AS72},
\beq\label{at}
A(t)=\sum_k[E_k(t)+D_k(t)]
=\sum_k {1\over2}a_k w(-u_k)\;.\label{w}
\eeq
(It is understood that $E_k(t)=0$ for poles in the lower half plane that 
have not been crossed when deforming the contour.) The second 
expression is very  
useful for studying the short time behaviour, 
but the first one has the advantadge of 
separating explicitly the exponential decay, $E_k$, from the 
``correction'' $D_k$, 
which is given in terms of the known entire function $w$
parameterized by the pole position and time.  Numerical values and
asymptotic properties of this function for small or large times are
easy to calculate. 

   The above treatment is easily extended for an $M(q)$ that includes 
an entire function in addition to the pole expansion.  
This would add to the $w$-functions the
integral along ${\cal D}$ of the entire function times a real Gaussian.

\subsection{Short time behaviour}
The short time behaviour of the quantum survival probability  is
easily analyzed in terms of the above formalism,  
which allows to classify 
several possible non exponential dependences. 

Many 
authors have described a short time $t^2$ dependence of the {\it decay
probability} $P_{\rm decay}\equiv 1-{\cal S}$ provided the mean energy and
second energy moment of these states exist, see in particular  
the work related to the ``quantum Zeno paradox'' \cite{MiSu}.
Less attention has been paid to the short time behaviour if these
conditions are not fulfilled. A formal treatment and
examples by Moshinsky and coworkers suggest a $t^{1/2}$ dependence of
the decay probability at short times
\cite{Moshinsky51,GLM}. We shall clarify how these two seemingly
different claims can be compatible, and describe other
possible dependences.

        The Taylor series (\ref{wser}) of the $w$ functions
	in (\ref{at}) gives a
series in powers of $t^{1/2}$,
\beq\label{A}
A(t)=\sum_k \frac{a_k}{2}\sum_{n=0}^{\infty}
\frac{[2^{-1}q_k(1-i)(t/m\hbar)^{1/2}]^n}{\Gamma({n\over2}+1)}.
\eeq
This suggests a short time $t^{1/2}$ dependence of the decay
probability, as claimed by Moshinsky and coworkers
\cite{Moshinsky51,GLM}.  On the other hand, the formal series based
on expanding the evolution operator,
\begin{equation}\label{texp}
A(t,\psi)=\langle \psi|e^{-iHt/\hbar}|\psi\rangle=
1-\frac{it}{\hbar}\langle \psi|H|\psi\rangle
-\frac{t^2}{2\hbar^2} \langle\psi|H^2|\psi\rangle+\cdots,
\end{equation}
provides a $t^2$ dependence, 
\beq
P_{\rm decay}=\frac{t^2}{\hbar^2}\bigl(\langle\psi|H^2|\psi\rangle-
\langle\psi|H|\psi\rangle^2\bigr)+\cdots.
\eeq
However, the expectation values of $H$ and/or higher powers of $H$
may not exist. Several behaviours are possible depending on the
existence of these moments. The question of the physical realizability 
of Hilbert space states with infinite first or second energy moments is
subject to debate \cite{Exner85}. 
We shall leave this debate aside, and determine the 
possible implications on the short time behavior.  
 
    Consider the first two derivatives of $A$ at time $t=0$
first from (\ref{texp}) and then by assuming a general short time
dependence of the form $A\sim 1+b\,t^c$, where $b$ and $c$ are finite
constants,
\beqa 
\frac{dA}{dt}\Big|_{t=0}&=&\frac{-i}{\hbar}\la \psi|H|\psi\ra=
b\,c\,t^{c-1}\big|_{t=0}\\
\frac{dA^2}{dt^2}\Big|_{t=0}&=&-\frac{1}{\hbar^2}\la\psi|H^2|\psi\ra
=b\,c\,(c-1)t^{c-2}\big|_{t=0}\,.
\eeqa
If the mean energy of the initial state does not exist, a $t^{1/2}$
dependence of the decay probability is possible, see examples 
in  \cite{MSW96} and \cite{GLM}. 

        If the mean energy is finite so that $d{\cal S}/dt|_{t=0}=2{\rm Re}
(dA/dt|_{t=0})=0$, then $c\geq 1$.  This rules out a $t^{1/2}$
dependence of $A$ since a $t^{1/2}$ dependence implies an infinite
time derivative of $A$ at $t=0$.  The corresponding coefficient for
$t^{1/2}$ in (\ref{A}) must vanish by compensation between the
different pole contributions.
  
      The second derivative is only finite at time zero if $c\ge 2$.
This means that if the first energy moment exists but not the second, 
a dependence $t^c$ where $1\le c<2$ is possible for $A$ (and for the
decay probability), in particular $t^{3/2}$.  (The coefficient for
$t^{1/2}$ must also vanish in this case.)  Otherwise, one can expect
that the series (\ref{texp}) will be effective at short times for states
with finite moments $\langle\psi|H^n|\psi\rangle$ leading to a $t^2$
behaviour. Examples where $t^{3/2}$ and $t^{2}$
dominate the short time behaviour of $P_{\rm decay}$ are provided 
in \cite{MSW96}.

\subsection{Large time behaviour\label{ltb}}
At first sight the asymptotic expansion of the $w$-function 
for $t\sim \infty$ in the correction term to 
the exponential decay suggests a long time dependence 
of the survival probability as $t^{-1}$, but in fact the 
general behaviour is $t^{-3}$ because of the cancellation of 
all the $t^{-1}$ contributions. 
Due to
the exponential $e^{-izt/\hbar}$ in (\ref{prop}) the large $t$
behaviour is dominated by the region around the origin. 
The origin is actually a
saddle point for the steepest descent path for this exponential factor
that crosses the origin along the diagonal ${\cal D}$ of the
second and fourth
quadrants. By introducing $u$ and $f$ variables as in (\ref{uf}) 
the exponential becomes $e^{-u^2}$ and $u$ remains real along
the steepest descent path. 
 
       The resolvent matrix element
$\langle \psi|(z-H)^{-1}|\psi \rangle$
which is defined for ${\rm Im} q>0$ (first energy sheet) has to be
analytically continued into the lower half $q$-plane (or second sheet
of the complex $z$ plane) to allow for this type of analysis,  
which will be valid in particular for finite range potentials. Provided
that the analytically continued function is analytical at the origin
it has a Taylor series expansion
\begin{equation}\label{gfs}
\langle \psi|(z-H)^{-1}|\psi\rangle=a_0+ a_1 q+ a_2 q^2+...
\end{equation}
with coefficients $a_i$ depending on $\psi$. But because of
the (odd) $q$ factor in (\ref{Iq}), the first term, $a_0$, does not
contribute to the integral (\ref{prop}). The asymptotic formula for
the survival amplitude comes
therefore from the second term and takes the form
\begin{equation}
\langle \psi |e^{-iHt/\hbar}|\psi\rangle \sim
\frac{i}{2m\pi}a_1 f^3 \intf du\, u^2 e^{-u^2}\\
=\frac{1-i}{2m\sqrt{\pi}} a_1 \,\left(\frac{m\hbar}{t}\right)^{3/2}\;.
\end{equation}
This formal result depends on the validity of (\ref{gfs}), and on the
assumption that no additional contributions due to the deformation of
the contour are to be considered asymptotically. In general the
analytically continued matrix elements of the resolvent will have
poles in the lower half $q$-plane that may be crossed when deforming
the contour, but these can only yield
contributions that decay {\it exponentially} with time, so they are
negligible at long times.

A similar analysis may be performed for the propagator 
(no bound states) \cite{MDS95}
\begin{eqnarray}\label{prop}
\langle x|e^{-iHt/\hbar}|x'\rangle&=&\frac{i}{2\pi}\int_{\cal C} dq\,
I(q) e^{-izt/\hbar}\;,
\\
\label{Iq'}
I(q)&=&{q\over m}\langle x|\frac{1}{z-H}|x'\rangle\;,
\end{eqnarray}
substituting  $M(q)$ by $I(q)$. 
Quite generally, $I(q)$ vanishes at
$q=0$, and a $t^{3/2}$ dependence results. 
An exception is free motion on the full line, 
where 
\begin{equation}\label{gf}
\langle x|{1\over z-H_0}|x'\rangle=\frac{-im}{q\hbar} e^{i|x-x'
|q/\hbar},
\end{equation}
so that $I(0)=-i/\hbar\ne 0$.  As a consequence, the
asymptotic behaviour of the probability density for free motion on the
full line is generically $t^{-1}$.
This is an important case in which
(\ref{gfs}) is {\it not} satisfied.  Explicitly, by carrying out
the integral in (\ref{prop}), the well known propagator
\begin{equation}
\langle x|e^{-iH_0t/\hbar}|x'\rangle=
\left(\frac{m}{iht}\right)^{1/2}e^{im(x-x')^2/2\hbar t}
\label{fmp}
\end{equation}
is obtained. 
A $t^{-1}$ behaviour will also occur exceptionally
when the potential allows for a zero energy pole of the 
resolvent.  

The free-motion probability density may decay faster than $t^{-1}$   
when the momentum amplitude $\la p|\psi\ra$ vanishes at $p=0$,
so that the $q^{-1}$ singularity is cancelled, see Figure 2. 
The exceptional cases of decay slower than $t^{-1}$ has been studied by 
Unnikrishnan \cite{Unnikrishnan98}. 

\begin{figure}[tbp]
\begin{center}
\includegraphics[width=.6\textwidth]{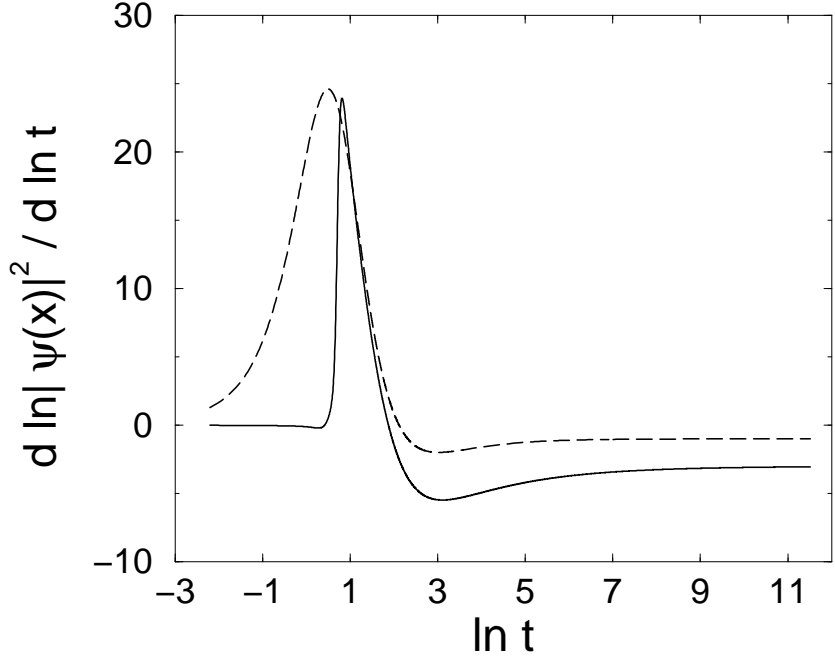}
\end{center}
\caption[]{$d \ln |\la x |\psi(t)\ra|^2/d \ln t$ versus $d\ln t$ 
for two different wave packets: one of them vanishes at $p=0$,  
$\la p|\psi(0)\ra=C(1-e^{-\alpha p^2/\hbar^2})e^{-\delta^2(p-p_0)^2/\hbar^2
-ipx_0/\hbar}\Theta(p)$ ({\it solid line}), and the other one is a 
a Gaussian wave packet,   
$\la p|\psi(0)\ra=C'e^{-\delta^2(p-p_0)^2/\hbar^2
-ipx_0/\hbar}$ ({\it dashed line}). $C$ and $C'$ are normalization 
constants; the parameters are     
$p_0=1$, $x_0=-10$, $\alpha=0.5$, $\delta=1$, $x=0$, $m=1$ 
(all quantities in atomic units).
Note the asymptotic dependences of the probability densities:   
$t^{-3}$ and $t^{-1}$ respectively.
}    
\label{asym}
\end{figure}

\section{Other characteristic times of wave propagation}

In the previous section we have seen how contour deformation techniques 
in the complex plane  allow to 
single out contributions to the survival amplitude from resonance poles.
In general the integral that provides the time dependent wave function 
may involve other critical points, ``structural'' poles,
saddle points, or branch points, that determine the transient and
asymptotic behaviour of the wave 
propagation. It is frequently possible to write explicit expressions or 
asymptotic expansions for 
the contributions of these critical points. In simple cases the effect of
(the dominant term of) one of the critical points  
provides already a good approximation and a simple picture emerges,
where characteristic times or velocities for the arrival of the main
signal may be identified.
Also typical is the transition from the dominance of one critical point to 
another, which may lead to a change in qualitative behaviour and
to a characteristic time for the transition.
The pioneering work in this
direction is due to
Stevens \cite{Stevens80,Stevens83}, who followed the techniques that  
Sommerfeld and Brillouin introduced in their study of the
propagation of light in dispersive media \cite{Brillouin60}. 
Examples of application to quantum scattering 
off a square barrier and a separable potential may be found in \cite{BM96}
and \cite{MP98JPA}. 
Here we shall examine, following \cite{MB00},
the somewhat simplified case corresponding to 
a point source producing evanescent waves following \cite{MB00}. This is  
not a ``scattering problem'' in the standard sense,
but it illustrates quite clearly the techniques and
concepts involved in more conventional scattering problems, and 
in other time dependent phenomena where a stationary 
state is achieved after a transient behaviour. 

In order to summarize essential aspects of the time dependence of wave
phenomena other characteristic velocities or times have been
traditionally defined. (We'll see that some of them coincide 
with times associated with critical points.) The
{\em phase velocity}, $\om/k$, is the velocity of constant phase
points in the stationary wave (assume $k>0$ for the time being)
\beq
e^{ikx-i\om t}\,.
\eeq
The boundary conditions, the superposition principle and the {\em
dispersion relation} $\om =\om (k)$ between the frequency $\omega$ and
the wavenumber $k$ determine the time evolution of the waves in a
given medium.  If a ``group'' is formed by superposition of stationary
waves around a particular $\omega$, it propagates with the {\em group
velocity} $d\om/dk$.  In {\em dispersive media} (where $\om$ depends
on $k$), the group velocity can be smaller (normal dispersion) or
greater (anomalous dispersion) than the phase velocity.  It was soon
understood that these velocities could be both greater than $c$ for
the propagation of light; Sommerfeld and Brillouin \cite{Brillouin60},
studying the fields that result from an input step function modulated
signal in a single Lorentz resonance medium, introduced other useful
velocities, such as the velocity of the very first wavefront (equal to
$c$), or the {\em signal velocity} for the propagation of the main
front of the wave.

The above description is however problematic for {\em evanescent
waves}, characterized by imaginary wavenumbers instead of the real
wavenumbers of propagating waves.  The role played by the imaginary
part of the group velocity $d\om /dk$ and the possible definition of a
signal velocity in the evanescent case have been much discussed.
Assume that a source is placed at $x=0$ and emits with frequency $\omega_0$
from $t=0$ on.
If $\o0$ is above the {\em cutoff frequency}
of the 
medium (the one that makes $k=0$)
a somewhat distorted but recognizable front propagates with
the velocity corresponding to $\o0$. 
For the dimensionless Schr\"odinger equation 
\beq
i\frac{\partial{\psi}}{\partial t}=
-\frac{\partial^2\psi}{\partial x^2}+\psi\,.
\label{dse}
\eeq
the dispersion relation takes the form 
\beq\label{disp}
\om=1+k^2\,, 
\eeq
and the signal propagation velocity for the main front is
equal to the group velocity,     
$v_p=(d\om/dk)_{\o0}=2(\o0-1)^{1/2}$. In other words, at some
distance $x$ form the source, the amplitude behaves, 
in first approximation, as
\beq
\psi(x,t)\approx e^{-i\o0 t}e^{+i k_0 x}\Theta(t-xv_p)\,
\eeq 
where $k_0=(\om_0-1)^{1/2}$ is the wavenumber related to  
$\o0$ by the dispersion relation, and $\Theta$ is the Heaviside (step) 
function.     
In the evanescent case, $\o0<1$, a preliminary analysis by Stevens 
\cite{Stevens80,Stevens83,Moretti92},
following the contour deformation techniques 
used  by Brillouin and Sommerfeld
suggested that a main front,
moving now with velocity $v_m=2(1-\o0)^{1/2}={\rm Im}(d\om/dk)_{\o0}$,
and attenuated exponentially 
by $\exp(\kappa_0 x)$ (where $\kappa_0=(1-\o0)^{1/2}$),
could be also identified,  
\beq\label{front}
\psi(x,t)\approx e^{-i\o0 t}e^{-\kappa_0 x}\Theta(t-xv_m)\,.
\eeq 
The contour for the integral defining the 
field evolution was deformed along the
steepest descent path from the saddle point; and  
the main front (\ref{front}) was associated with  a residue due to
the crossing of a pole at $i\kappa_0$ by the
steepest descent path.

The result seemed to be supported by a different approximate 
analysis of Moretti based on the exact solution \cite{Moretti92},
and by the fact that    
the time of arrival of the evanescent front, 
$\tau=x/v_m$, had been found independently by B\"uttiker and Landauer
\cite{BL82,Buttiker83} as a characteristic {\em traversal time}
for tunnelling
using rather different criteria (semiclassical arguments, 
the rotation of the electron spin in a weak magnetic field,
and the transition from 
adiabatic to sudden regimes in an oscillating potential barrier).

However, more accurate studies of the point source problem and
other boundary conditions have shown that the contribution from the saddle
point (due to  frequency components above or at the frequency cutoff
created by the sharp onset of the source emission), and possibly from
other critical points (e.g.  resonance poles when a square barrier is
located in front of the source \cite{BM96}) are generally dominant at
$\tau$, so that no sign of the $\omega_0-$front is seen in the
total wave density at that
instant, see  \cite{RMFP90,RMA91,TKF87,JJ89,BM96,MB00}, and the subsection
\ref{rtt} below.

Buttiker and Thomas  
reconsidered the signal sent out by a source which has a sharp
onset in time \cite{BT98}. 
They proposed two approaches to enhance the monochromatic fronts 
compared to the forerunners due to the saddle.
First, the dominance of 
the high frequencies forerunners
could be avoided if the 
source is frequency limited such that all frequencies of the source 
are within the evanescent case. Of course this makes the 
onset of the signal unsharp.   
A second option is not to limit the 
source but to frequency limit the detection. We can chose a detector 
that is tuned to the frequency of the source and that responds when 
the monochromatic front arrives.

These two proposals and the sharp onset case 
were later implemented
and examined in detail by Muga and Buttiker \cite{MB00}. 
For a source with a sharp onset,   
they found that the traversal time $\tau$ plays a basic and  
unexpected role in the transient regime. 
For strongly attenuating conditions 
(in the WKB-limit) the traversal time governs the appearance 
of the first main peak of the forerunner. In contrast, 
the transition from the forerunner to an asymptotic regime 
which is dominated by the monochromatic signal of the source 
is given by an exponentially long time, see more details in 
\ref{rtt} below.
If the source is frequency band limited such that it switches on gradually
but still fast compared to the traversal time, the situation 
remains much the same as for the sharp source, except that 
now the transition fom the transient regime to the stationary 
regime occurs much faster, but still on an exponentially long time-scale. 
The situation changes if we permit the source to be switched  
on a time scale comparable to or larger than the traversal time for
tunneling. Clearly, in this case a precise definition of the traversal 
time is not possible. But for such a source the transition 
from the transient regime to the asymptotic regime 
is now determined by the traversal time. 
Much the same picture emerges if we limit instead of the 
source the detector. Muga and Buttiker modell the 
detector response by means of a ``spectrogram'',
a time-frequency representation 
of the wave function at a fixed point. 
As long as the frequency window of the detector 
is made sharp enough to determine the traversal time with accuracy, 
the detector response is dominated by the uppermost frequencies. 
In contrast, if the frequency window of the detector is made so narrow 
that the possible uncertainty in the determination of the traversal 
time is of the order of the traversal time itself, the detector 
sees a crossover from the transient regime to the monochromatic 
asymtotic regime at a time determined by the traversal time. 

Possibly, the fact that we can not determine the traversal time 
with an accuracy  better than the traversal time itself
tells us something fundamental about the tunneling time problem 
and is not a property of the two particular methods investigated.

\subsection{Role of the traversal time for a source with a sharp
onset\label{rtt}}
We shall obtain exact and approximate expressions of the time dependent
wave function for $x>0$ and $t>0$ corresponding to the 
Schr\"odinger equation (\ref{dse}) and the ``source boundary condition''
\beq\label{initt}
\psi(x=0,t)=e^{-i\o0 t}\Theta(t)\,,
\eeq
in the evanescent case $\o0<1$.  
(A discussion of the physical meaning of ``source boundary conditions''
as compared to standard ``initial value'' conditions has been 
presented recently \cite{BEM01JPA}.) 
The solution may be constructed from its 
Fourier transform 
as  
\beq
\label{two}
\psi(x,t)=
-\frac{e^{-it}}{2\pi i}\int_{\Gamma_+} dk
\left[\frac{1}{k+i\kappa_0}+\frac{1}{k-i\kappa_0}\right]
e^{ikx-ik^2t}\,,
\eeq
where the contour $\Gamma_+$ goes from $-\infty$ to $\infty$
passing above the pole at $i\kappa_0$, 
and 
\beq
\kappa_0=(1-\o0)^{1/2}\;.
\eeq
The contour can be deformed   
along the steepest descent path from the saddle at $k_s=x/2t$, 
the straight line
\beq
k_I=-k_R+x/2t\, ,
\eeq
($k_R$ and $k_I$ are the real and imaginary parts of $k$.) 
plus a small circle around 
the pole
at $i\kappa_0$ after it has been crossed by the steepest descent path, 
for fixed $x$, at the 
critical time  
\beq
\tau=\frac{x}{2\kappa_0}\,. 
\eeq

This procedure allows to  
recognize two $w$-functions \cite{AS72}
one for each integral,   
\beq\label{exact}
\psi(x,t)=\frac{1}{2}e^{-it+ik_s^2 t}\left[
w(-u_0')+w(-u_0'')\right]\,.
\eeq
Here,  
\beqa\label{us}
u_0'&=&\frac{1+i}{2^{1/2}}t^{1/2}\kappa_0\left(-i-\frac{\tau}{t}\right)
\\
\nonumber
u_0''&=&\frac{1+i}{2^{1/2}}t^{1/2}\kappa_0\left(i-\frac{\tau}{t}\right)\,.
\eeqa
It is clear from the exact result (\ref{exact},\ref{us}),
that $\tau$ is an important parameter that appears naturally in
the $w$-function arguments, and determines with $\kappa_0$
the global properties of the solution. Its detailed role will be discussed 
next.    

The simplest approximation for $\psi(x,t)$ for times before $\tau$
is to retain the
dominant contribution of the saddle
by putting $k=k_s$ in the denominators of 
(\ref{two}) and integrating along the steepest descent path,  
\beq
\psi_s(x,t)=\frac{e^{-it+ik_s^2t}}{2i\pi^{1/2}}\left(\frac{1}{u_0'}
+\frac{1}{u_0''}\right)
\label{psis}
\eeq
The average local instantaneous frequency for this saddle 
contribution is equal to the frequency of the saddle 
point \cite{MB00},   
\beq\label{ws}
\os\equiv 1+x^2/4t^2\,. 
\eeq
After the crossing of the pole $i\kappa_0$ by the steepest descent 
path at $t=\tau$ the residue 
\beq\label{resi}
\psi_0(x,t)=e^{-i\o0 t}e^{-\kappa_0 x}\Theta(t-\tau) .
\eeq 
has to be added to (\ref{psis}), 
\beq\label{apro}
\psi(x,t)\approx \psi_s(x,t)+\psi_0(x,t)\,.
\eeq
The solution given by Eq. (\ref{resi}) describes a monochromatic front 
which carries the signal into the evanescent medium. 
The conditions of validity of this  
approximation can be determined by examining the asymptotic
series of the $w(z)$ functions in (\ref{exact}) for large $|z|$, 
see the Appendix. In fact (\ref{apro}) is obtained from the dominant
terms of these expansions. Large values of $|z|$ are obtained with     
large values of $\kappa_0$, $t$, or
$x$, and also when $t\to 0$. 
Within the conditions that make the saddle approximation valid, the
contribution of the pole is negligible. 
To see this more precisely 
let us examine the ratio between the 
modulus of the two contributions,  
\beq
R(t)\equiv\frac{|\psi_0|}{|\psi_s|}=
\frac{2\pi^{1/2}}{x} e^{-\kappa_0 x}t^{3/2}(x^2/4t^2+\kappa_0^2)\,.
\eeq
Its value at $\tau$ is an exponentially small quantity, 
\beq\label{Rt0}
R(t=\tau)=e^{-\kappa_0 x}(2\pi\kappa_0 x)^{1/2}\,.
\eeq
In summary, for the source 
with a sharp onset described here, the monochromatic front is not 
visible when the approximation 
(\ref{apro}) remains valid around $t=\tau$. 
A complementary analysis is carried out in Chapter ....   
 
However two very important observable features of the wave can be
extracted easily from (\ref{apro}). 
The first one is the arrival of the {\em transient front}, characterized 
by its maximum density at $t_f\equiv\tau/3^{1/2}$.
This
time is of the order of $\tau$, but the wave front that arrives does not 
oscillate with the pole frequency $\o0$, but with the saddle point 
frequency $\os$. 

The second observable feature that we can extract from (\ref{apro})
is the time scale for the 
attainment of the stationary regime, or equivalently, the duration 
$t_{tr}$ of the transient regime dominated by the saddle before the 
pole dominates. $t_{tr}$ can be identified formally as the time where
the saddle and pole contributions
are equal, $R=1$. Because of 
(\ref{Rt0}) we shall assume  $\tau<<t_{tr}$ to obtain
the explicit result 
\beq\label{ttr}
t_{tr}\approx
\left(\frac{x e^{\kappa_0 x}}{2\kappa_0^2\pi^{1/2}}\right)^{2/3}\,.
\eeq
Finally, when $x\kappa_0$ is small ($\stackrel{<}{\sim}1$),
the saddle approximation  describes correctly 
the very short time initial growth, but fails around $\tau$  
because the pole is within the width of the Gaussian
centered at the saddle point. The pole cancels part of the 
Gaussian contribution so that the bump predicted by 
$\psi_s$ at $\tau/3^{1/2}$ is not seen in this regime. 
$\tau$ does not correspond to any 
sharply defined feature, but provides a valid rough estimate of the 
attainment of the stationary regime.

\subsection*{Acknowledgements}This chapter
is largely based on work done in collaboration
with S. Brouard, J. P. Palao, and R. Sala in La Laguna,   
G. Wei and R. F. Snider in Vancouver, and M. Buttiker in Geneve.
Thanks to them all, and    
to I. L. Egusquiza for commenting on the manuscript.
I acknowledge support by the Basque Government (PI-1999-28), 
and MCYT (BFM2000-0816-C03-03).

\appendix

\section*{Appendix: Properties of $\lowercase{w}$-functions}
The $w-$function is an entire function defined in terms of
the complementary error function as \cite{AS72}
\beq
w(z)=e^{-z^2}{\rm erfc}(-iz)\,.
\eeq
$w(z)$ is frequently recognized by its integral expression  
\beq\label{wint}
w(z)=\frac{1}{i\pi}\int_{\Gamma_-}\frac{e^{-u^2}}{u-z}du
\eeq
where $\Gamma_-$ goes from $-\infty$ to $\infty$ passing below 
the pole at $z$. For ${\rm Im} z>0$ this corresponds to an integral
along the 
real axis. For ${\rm Im} z<0$ the contribution of the residue has to be
added, and for ${\rm Im} z=0$ the integral becomes the principal part 
contribution along the real axis plus half the residue. 
From (\ref{wint}) two important properties are deduced, 
\beqa\label{A3}
w(-z)&=&2e^{-z^2}-w(z)\\
\noalign{\hbox{\rm and}}
w(z^*)&=&[w(-z)]^*.
\eeqa
To obtain an asymptotic series 
as $z\to\infty$ for ${\rm Im} z>0$ one may expand  
$(u-z)^{-1}$ around the origin (the radius
of convergence is the distance
from the origin to the pole, $|z|$) and integrate term by term. This
provides   
\begin{equation}\label{was}
w(z)\sim \frac{i}{\sqrt{\pi}\, z}\left[1+\sum_{m=1}^\infty
\frac{1\cdot 3\cdot ...\cdot (2m-1)}{\left(2z^2\right)^m}\right]
\,\;
{\rm Im} z>0
\end{equation}
which is a uniform expansion in the sector ${\rm Im} z>0$.
For the sector ${\rm Im} z<0$ (\ref{A3}) gives
\beq\label{zle}
w(z)\sim \frac{i}{\sqrt{\pi}\, z}\left[1+\sum_{m=1}^\infty
\frac{1\cdot 3\cdot ...\cdot (2m-1)}{\left(2z^2\right)^m}\right]
+2 e^{-z^2}  
\,,\;
{\rm Im} z<0\,.
\eeq
If $z$ is in one of the bisectors then $-z^2$ is purely imaginary 
and the exponential becomes dominant.
But right at the crossing of the real axis, ${\rm{Im}} z=0$,
the exponential term is of 
order $o (z^{-n})$, (all $n$), so that (\ref{was}) and (\ref{zle}) are
asymptotically equivalent as $|z|\to\infty$.

$w(z)$ has the
series expansion
\begin{equation}\label{wser}
w(z)=\sum_0^\infty\frac{(iz)^n}{\Gamma(\frac{n}{2}+1)}.
\end{equation}
The $w$-function is a particular case of the Moshinsky
function \cite{Moshinsky51}, which can be regarded as ``the
basic propagator
for a Schr\"odinger transient mode''
\cite{Nussenzveig92}.

\end{document}